\documentclass[journal]{IEEEtran}

\usepackage{graphicx}
\usepackage{float}
\usepackage{mathrsfs}
\usepackage{amsfonts}
\usepackage{color}
\usepackage{multirow}
\usepackage{cite}
\usepackage{caption}
\usepackage{graphicx} 
\usepackage{epstopdf}

\usepackage{algorithm}  
\usepackage{algorithmicx}  
\usepackage{algpseudocode}  
\usepackage{amsmath}
\usepackage{diagbox}

\ifCLASSINFOpdf

\else

\fi

\begin{document}

\title{When Blockchain Meets AI: Optimal Mining Strategy Achieved By Machine Learning}


\author{\IEEEauthorblockN{Taotao Wang, Soung Chang Liew, and Shengli Zhang}
\thanks{T. Wang and S. Zhang are the College of Electronics and Information Engineering, Shenzhen University, Shenzhen 518060, China (e-mail: ttwang@szu.edu.cn; zsl@szu.edu.cn). 
	
S. Liew is with the Department of Information Engineering, The Chinese University of Hong Kong, Hong Kong SAR, China (e-mail: soung@ie.cuhk.edu.hk)}

\thanks{This research was funded by the National Key R\&D Program of China (2018YFB2100705) and the Natural Science Fund of Guangdong Province (2020A1515010708).}

}

\markboth{This work was accepted for publication in International Journal of Intelligent Systems}%
{Shell \MakeLowercase{\textit{et al.}}: Bare Demo of IEEEtran.cls for Journals}

\maketitle

\begin{abstract}
This work applies reinforcement learning (RL) from the AI machine learning field to derive an optimal Bitcoin-like blockchain mining strategy. A salient feature of the RL learning framework is that  an optimal (or near optimal) strategy can be obtained without the knowing the details of the blockchain network model.  Previously, the most profitable mining strategy was believed to be honest mining encoded in the default blockchain protocol. It was shown later that it is possible to gain more mining rewards by deviating from honest mining. In particular, the mining problem can be formulated as a Markov Decision Process (MDP) which can be solved to give the optimal mining strategy. However, solving the mining MDP requires knowing the values of various parameters that characterize the blockchain network model. In real blockchain networks, these parameter values are not easy to obtain and may change over time. This hinders the use of the MDP model-based solution. In this work, we employ RL to dynamically learn a mining strategy with performance approaching that of the optimal mining strategy. Since the mining MDP problem has a non-linear objective function (rather than linear functions of standard MDP problems), we design a new multi-dimensional RL algorithm to solve the problem. Experimental results indicate that, without knowing the parameter values of the mining MDP model, our multi-dimensional RL mining algorithm can still achieve optimal performance over time-varying blockchain networks.
\end{abstract}


\IEEEpeerreviewmaketitle

\section{Introduction}
\IEEEPARstart{T}{HE} early digital cryptocurrencies rely on central authorities to settle transactions. Digital cryptocurrencies did not flourish, until the advent of Bitcoin \cite{nakamoto2008bitcoin, antonopoulos2014mastering}. To avoid single points of failure, Bitcoin is designed as a decentralized system without a central authority that could be compromised by corruption and attacks \cite{nakamoto2008bitcoin}. Since the birth of Bitcoin in 2008, it has become a widely accepted currency all over the world. In early 2018, the market price of Bitcoin went as high as 20,000 US dollars, reflecting robust demands and enthusiasm for Bitcoin by the public. 

The security of Bitcoin is built on the foundation technology of blockchain. Blockchain contains several key technical components, including its chained data structure, peer-to-peer network protocol, and distributed consensus algorithm \cite{tschorsch2016bitcoin,consensus2019,puthal2018everything}.  Blockchain has become a cutting-edge technology in FinTech \cite{fanning2016blockchain}, Internet of Things (IoT) \cite{ferrag2018blockchain, dai2019blockchain}, and supply chains \cite{abeyratne2016blockchain}. The Bitcoin’s blockchain is not controlled by a central authority; it is assembled by peers in the network independently in a distributed manner. In order that the blockchains maintained by different peers are consistent, the peers must agree on a single universal truth about the transactions of Bitcoin through a consensus-building process.

Consensus in the Bitcoin network is achieved by the proof-of-work (PoW) consensus algorithm. The idea of PoW originated in \cite{dwork1992pricing} and is rediscovered and exploited in the implementation of Bitcoin. PoW provides strong probabilistic consensus guarantee with resilience against up to 1/2 malicious nodes \cite{garay2015bitcoin,pass2017analysis}. The successful operation of Bitcoin demonstrates the practicality of using PoW to achieve consensus. Subsequent to Bitcoin, many other cryptocurrencies, such as Litecoin \cite{lee2011litecoin}, Ethereum \cite{buterin2014ethereum}, also adopt the PoW consensus algorithm.

Peers running the PoW consensus algorithm are miners who compete to solve a difficult cryptographic hash puzzle, called the PoW problem. The miner who successfully solves the PoW problem obtains the right to extend the blockchain with a block consisting of valid transactions. In doing so, the miner receives a reward in the form of a newly minted coin written into the added block. Solving the PoW problem for rewards is called mining, just like mining for precious metals. 

Miners commit computation resources to solve the PoW problem. Previously, it was believed that the most profitable mining strategy is honest mining, wherein a miner will broadcast the newly added block as soon as it has solved the PoW problem. Let $\alpha$ be the ratio of a particular miner's computing power over the computing powers of all miners. This ratio is also the probability that the miner can solve the PoW problem before others in each round of an added block \cite{tschorsch2016bitcoin}. Over the long term, the rewards to a miner that executes the honest mining strategy are therefore $\alpha$ fraction of the total rewards issued by the Bitcoin network. This is reasonable since miners share the pie in proportion to their investments. Not known were whether there are other mining strategies more profitable than honest mining. 

Later, the authors of \cite{eyal2018majority} developed a selfish mining strategy that can earn higher rewards than honest mining. A selfish miner does not broadcast its mined block immediately; it carries out a block-withholding attack by secretly linking its future mined blocks to the withheld mined block. If the selfish miner can mine two successive blocks before other miners do, it can broadcast its two blocks at the same time to override the block mined by others. Since Bitcoin has an inherent self-adjusting mechanism to ensure that on average only one block is added to the blockchain every 10 minutes \cite{kraft2016difficulty}, by invalidating the blocks of others (hence, removing them from the blockchain), the selfish miner can increase its own profits. For example, with computing power ratio $\alpha  = {1 \mathord{\left/
		{\vphantom {1 4}} \right.
		\kern-\nulldelimiterspace} 4}$, the rewards obtained by selfish mining can be up to $1/3$ fraction of the total rewards  \cite{eyal2018majority}.  Based on this observation, \cite{nayak2016stubborn} further proposed various selfish mining strategies with even higher rewards. Despite the many versions of selfish mining, the optimal (i.e., most-profitable) mining strategy remained elusive until \cite{sapirshtein2016optimal}. 

The authors of \cite{sapirshtein2016optimal} formulated the mining problem as a general Markov Decision Process (MDP) with a large state-action space. The objective of the mining MDP, however,  is not a linear function of the rewards as in standard MDPs. Thus, the mining MDP cannot be solved using a standard MDP solver. To solve the problem, \cite{sapirshtein2016optimal} first transformed the mining MDP with the non-linear objective to a family of MDPs with linear objectives, and then employed a standard MDP solver over the family of MDPs to iteratively search for the optimal mining strategy. 

The approach in \cite{sapirshtein2016optimal} is model-based in that various parameter values (e.g., $\alpha$) must be known before the MDP can be set up. In real blockchain networks, the exact parameter values are not easy to obtain and may change over time, hindering the practical adoption of the solution. In this paper, we propose a model-free approach that solves the mining MDP using machine learning tools. In particular, we solve the mining MDP using reinforcement learning (RL) without the need to know the parameter values in the mining MDP model. 

RL is a machine-learning paradigm, where agents learn successful strategies that yield the largest long-term reward from trial-and-error interactions with their environment \cite{sutton2018reinforcement, kaelbling1996reinforcement}. Q-learning is the most popular RL technique \cite{watkins1992q}. It can learn a good policy by updating a state-action value function without an operating model of the environment. RL has been successfully applied in many challenging tasks, e,g., playing video games \cite{mnih2015human} and Go  \cite{silver2017mastering}, and controlling robotic movements \cite{schulman2015trust}. 

The original RL algorithm cannot deal with the nonlinear objective function of our mining problem. In this paper, we put forth a new multi-dimensional RL algorithm to tackle the problem. Experimental results indicate that our multi-dimensional RL mining algorithm can successfully find the optimal strategy. Importantly, it demonstrates robustness and adaptability to a changing environment (i.e., parameter values changing dynamically over time).

\section{Blockchain Preliminaries}

Blockchain is a decentralized append-only ledger for digital assets. The data of blockchain is replicated and shared among all participants. Its past recorded data are tamper-resistant and participants can only append new data to the tail-end of the chain of blocks. The state of blockchain is changed according to transactions, and transactions are group into blocks that are appended to the blockchain. The header of the block encapsulates the hash of the preceding block, the hash of this block, the Merkle root\footnote{The Merkle root of the transactions is the hash value of the Merkle tree whose leaves are the transactions \cite{merkle1987digital}.} of all transactions contained in this block, and a number called nonce that is generated by PoW. Since each block must refer to its preceding block by placing the hash of its preceding block in its header, all the blocks form a chain of blocks arranged in chronological order. Fig. 1 illustrates the data structure of blockchain.

\begin{figure*}[t]
	\centering
	\includegraphics[width=5.5in]{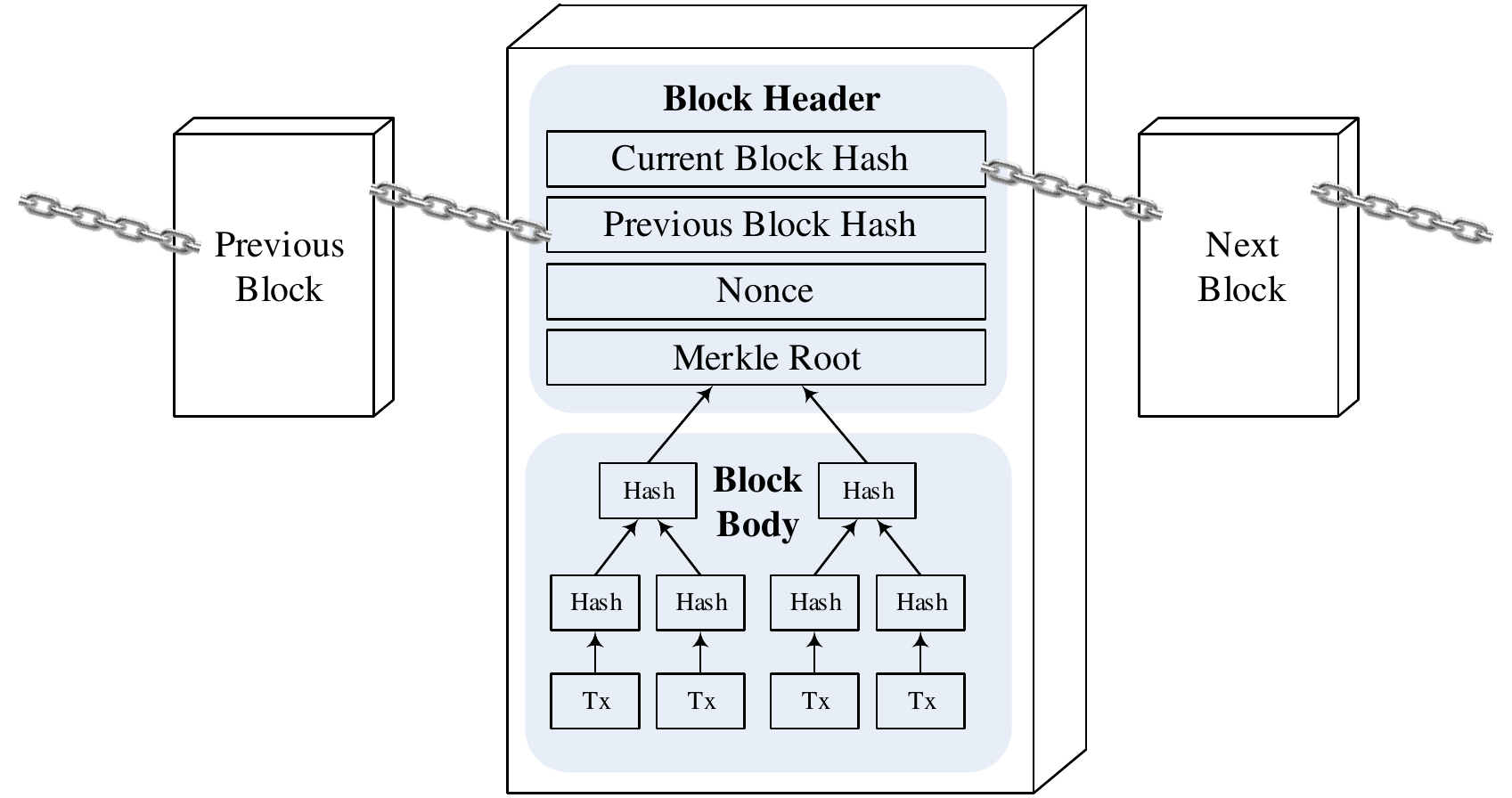}
	\caption{Data structure of blockchain.} \label{fig1}
\end{figure*}

\subsection{Proof of Work and Mining}

In this paper, we focus on a Bitcoin-like blockchain that adopts the PoW consensus protocol to validate new blocks in a decentralized manner.\footnote{There are also blockchains adopting other several consensus algorithms, such as Proof of Stake (PoS), and Byzantine fault tolerance (BFT) \cite{consensus2019}.} In each round, the PoW protocol selects a leader that is responsible for packing transactions into a block and appends this block to the blockchain. To prevent adversaries from monopolizing the blockchain, the leader selection must be approximately random. Since Bitcoin-like blockchain is permissionless and anonymity is inherently designed as the goal, it must consider the Sybil attack where an adversary simply creates many participants with different identities to increase its probability of being selected as the leader. To address the above issues, the key idea behind PoW is that a participant will be randomly selected as the leader of each round with a probability in proportion to its computing power.  

In particular, blockchain implements PoW using computational hash puzzles. To create a new block, the nonce placed into the header of the block must be a solution to the hash puzzle expressed by the following inequality  
\begin{equation}
{\cal H}\left( {n,p,m} \right) < D
\end{equation}
where the nonce $n$, the hash of the previous block $p$, the Merkle root of all included transactions $m$ are taken as the input of a cryptographic hash function ${\cal H}( \cdot )$ and the output of the hash function should fall below a target $D$ that is small with respect to the whole range of the hash function outputs. The used hash function (e.g., SHA-256 hash is used for Bitcoin) satisfies the property of puzzle friendliness \cite{wang2018research}: it is challenging to guess the nonce to fulfill (1) by a one-shot try. The only way to solve (1) is to try a large number of nonces one by one to check if (1) is fulfilled until one lucky nonce is found. Therefore, the probability of finding such a nonce is proportional to the computing power of the participant---the faster the hash function in (1) can be computed in each trial, the more nounces can be tried per unit time. Using the blockchain terminology, the process of computing hashes to find a nonce is called \emph{mining}, and the participants involved are called \emph{miners}.

\subsection{Honest Mining Strategy}

When a miner tries to append a new block to the latest legal block by placing the hash of the latest block in the header of the new block, we say that the miner mines on the latest block. The blockchain is maintained by miners in the following manner.

To encourage all miners to mine on, and maintain, the current blockchain, a \emph{reward} is given as an incentive to the miner by placing a coin-mint transaction in its mined block that credits the miner with some new coins. If the block is verified and accepted by other peers in the blockchain network, the reward is effective and thus can be spent on the blockchain. When a miner has found an eligible nonce, it publishes his block to the whole blockchain network. Other miners will verify the nonce and transactions contained in that block. If the verification of the block is passed, other miners will mine on the block (implicitly accepting the block); otherwise, other miners discard the block and will continue to mine on the previous legal block. 

If two miners publish two different legal blocks that refer to the same preceding block at the same time, the blockchain is then forked into two branches. This is called \emph{forking} of blockchain. Forks in the blockchain because they are manifestations of disagreement among peers on the blockchain structure. It can also compromise the integrity and security of the blockchain \cite{bagaria2018deconstructing}. To resolve a fork, PoW prescribes that only the rewards of the blocks on the longest branch (called the main chain) are effective. Then, miners are incentivized to mine on the longest branch, i.e., miners always add new blocks after the last block on the longest main chain that is observed from their local perspectives. If the forked branches are of equal length, miners may mine subsequent blocks on either branch randomly. This is referred to as the rule of the longest chain extension. Eventually, one branch will predominate and the other branches are discarded by peers in the blockchain network. 

The mining strategy adhering to the rule of the longest chain extension and publishing a block immediately after the block is mined is referred to as \emph{honest mining} \cite{tschorsch2016bitcoin,consensus2019,puthal2018everything}. The miners that comply with honest mining are called honest miners. It was widely believed that the most profitable mining strategy for miners is honest mining; and that when all miners adopt honest mining, each miner is rewarded in proportion to its computing power \cite{tschorsch2016bitcoin,consensus2019,puthal2018everything}. As a result, any rational miner will not deviate from honest mining.  This belief was later shown to be ill-founded and that other mining strategies with higher profits are possible \cite{eyal2018majority, nayak2016stubborn, sapirshtein2016optimal}. We will briefly discuss these mining strategies in the next section. For a more concrete exposition, we will first present the mining model.

\section{Blockchain Mining Model}

In this section, we present the Markov Decision Process (MDP) model for blockchain mining. Ref.  \cite{eyal2018majority} first developed an MDP mining model and used the model to construct a selfish mining strategy with higher rewards than honest mining. Then, \cite{nayak2016stubborn} proposed even more profitable selfish mining strategies. Recently, \cite{sapirshtein2016optimal} extended the MDP mining models of \cite{eyal2018majority, nayak2016stubborn} to a more general form. In this work, we adopt the mining model of \cite{sapirshtein2016optimal}. 

Without loss of generality, we assume the network is split into two mining pools: one is an adversary that controls a fraction $\alpha$  of the whole network's computing power; the other is the network of honest miners that controls a fraction  $1-\alpha $ of the computing power of the whole network. 

Even if the adversary and an honest miner release their newly mined blocks to the network simultaneously, the blocks will not be received by all miners simultaneously due to propagation delays and network connectivity. We model the communication capability of the adversary using the parameter $\gamma $, defined as the fraction of the honest miners that will first receive the block from the adversary when the adversary and one honest miner release their blocks approximately at a same time---more specifically, $\gamma (1 - \alpha )$ is the computing power of the honest network that will mine on the block of the adversary when the adversary and an honest miner release their blocks simultaneously. 

As in \cite{sapirshtein2016optimal}, we model blockchain mining as a single-player MDP $M = \left\langle {S,A,P,R} \right\rangle $, where $S$ is the state space, $A$ is the action space, $P$ is the transition probability matrix and $R$ is the reward matrix. Each transition is triggered by the event of a miner mining a new block, whether the block is mined by the adversary or one of the honest miners. The action taken by the adversary based on the previous state, together with the event, determines the next state to which the system evolves.

The objective of the adversary is to earn rewards higher than its computational power. To achieve this, the adversary will generally deviate from honest mining by building a private chain of blocks without releasing them the moment the blocks are mined; the adversary will release several  blocks from its private chain at a time to undo the honest chain opportunistically.

\begin{figure}[t]
	\centering
	\includegraphics[width=2in]{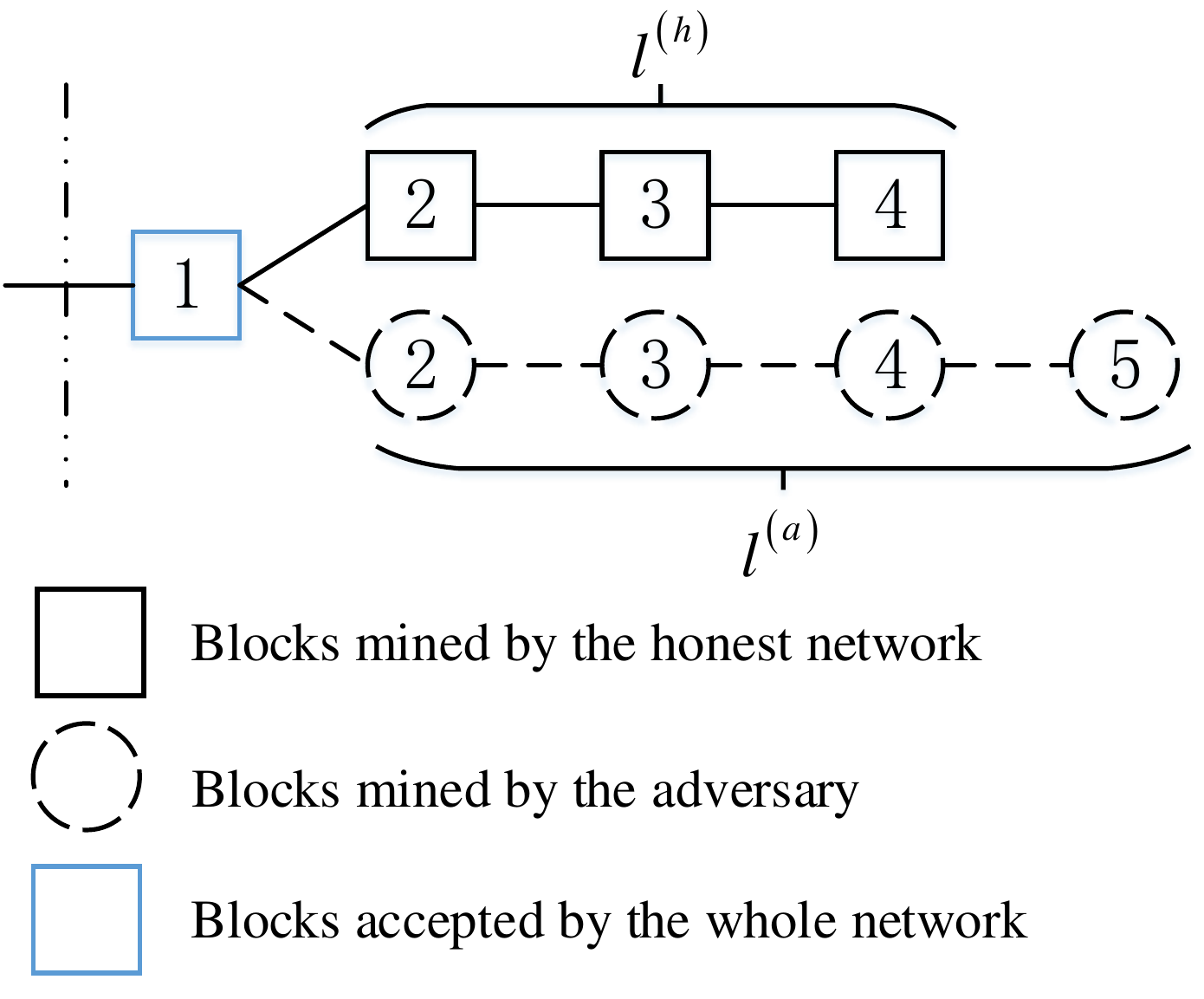}
	\caption{An illustrating example of the state in the adopted MDP.} \label{fig_bc_state}
\end{figure}

{\bf{State}}: Each state in the state space is represented by a three-tuple form $\left( {{l^{\left( a \right)}},{l^{\left( h \right)}},fork} \right)$, where ${l^{\left( a \right)}}$ and ${l^{\left( h \right)}}$  are respectively the lengths of the adversary’s chain and the honest network’s chain after the latest fork (as illustrated in Fig. \ref{fig_bc_state}). In general, $fork$ can take three possible values $\left( {irrelevant,relevant,active} \right)$. Their meanings will be explained later.

{\bf{Action}}: The action space $A$ includes four actions that can be executed by the adversary. 
\begin{itemize}
	\item  \emph{Adopt}: The adversary accepts the honest chain and mines on the last block of the honest chain. This action discards the ${l^{\left( a \right)}}$  blocks in the chain of the adversary and it renews the attack from the new starting point without a fork. This action is allowed by the MDP model for all ${l^{\left( a \right)}}$    and  ${l^{\left( h \right)}}$.
	
	\item	\emph{Override}: The adversary publishes one block more than the honest chain (i.e., ${l^{\left( h \right)}} + 1$ blocks) to the whole network. This action overrides the conflicting blocks of the honest chain. This action is allowed when ${l^{\left( a \right)}} > {l^{\left( h \right)}}$.

	\item \emph{Match}: The adversary publishes the same number of blocks as the honest chain (i.e., ${l^{\left( h \right)}}$  blocks) to the whole network. This action creates a fork deliberately and initiates an open mining competition between the two branches of the adversary and the honest network. This action is allowed when ${l^{\left( a \right)}} \ge {l^{\left( h \right)}}$ and $fork = relevant$.
	
	\item\emph{Wait}: The adversary does not publish blocks and it just keeps mining on its own chain. This action is always feasible.

\end{itemize}

One remark about the actions of the MDP mining model is that some actions that can generally be performed are deliberately removed from the action-state space because these actions are not gainful for the adversary. For example, when ${l^{\left( a \right)}} < {l^{\left( h \right)}}$, the adversary can still release a certain number of its blocks. However, since releasing fewer blocks than the number of blocks on the honest chain will not increase its probability of mining the next block compared to mining it privately, these actions thus are excluded from the allowed actions.


We now explain the three values of the entry $fork$ in the three-tuple state. 
\begin{itemize}
	\item  $Relevant$: The value of relevant means that the latest block is mined by the honest network. Now, if   $fork = relevant$ and ${l^{\left( a \right)}} \ge {l^{\left( h \right)}}$, the action $match$ is allowed. For example, if the previous state is $\left( {{l^{\left( a \right)}},{l^{\left( h \right)}} - 1, \bullet } \right)$
	  and now the honest network successfully mines one block, the state then changes to $\left( {{l^{\left( a \right)}},{l^{\left( h \right)}},relevant} \right)$. If at this time, ${l^{\left( a \right)}} \ge {l^{\left( h \right)}}$, $match$ is allowed. We remark that $match$ here may be gainful for the adversary because $\gamma (1 - \alpha )$  computing power of the honest network would be dedicated to mining on the adversary chain because of the near-simultaneous releases of the latest block of the adversary chain and the latest block of the honest chain.  In this state, as far as the public is concerned, there no fork yet, since the ${l^{\left( a \right)}}$ mined blocks of the adversary are private and hidden from the public. However, if the adversary execute a match from this state, then a fork will be made known to the public and an active competition between the two branches will follow.

	\item $Irrelevant$: The value of $irrelevant$ means that the latest block is mined by the adversary and the blocks published by the honest network have been already received by (the majority of) the honest network. Now, even if ${l^{\left( a \right)}} \ge {l^{\left( h \right)}}$, the action $match$ is not allowed. For example, if the previous state is $\left( {{l^{\left( a \right)}} - 1,{l^{\left( h \right)}}, \bullet } \right)$ and now the adversary successfully mines a new block, the state changes to $\left( {{l^{\left( a \right)}},{l^{\left( h \right)}},irrelevant} \right)$. We emphasize that $match$ is disallowed here even if ${l^{\left( a \right)}} \ge {l^{\left( h \right)}}$, not because it cannot be performed in the blockchain, but rather $match$ here is not gainful for the adversary. If $match$ were to be performed here, no computing power of the honest network would shift to mining on the adversary chain because the miners in the honest network would have received the latest block of the honest chain first (well before the current transition triggered by the adversary mining a new block) and would have dedicated to mining on the honest chain already. Again, in this state, there is no fork as far as the public blockchain is concerned.

	\item $Active$: The value of $active$ means that the adversary has executed the action $match$ from the previous state, and the blockchain is now split into two branches.  For example, if the previous state is $\left( {{l^{\left( a \right)}},{l^{\left( h \right)}},relevant} \right)$ with ${l^{\left( a \right)}} \ge {l^{\left( h \right)}}$ and the adversary executed the action $match$. If the new transition is triggered by the honest network mining a new block, then the state transitions to $\left( {{l^{\left( a \right)}} - {l^{\left( h \right)}},1,active} \right)$. In short, $active$ means a fork is made known to the public and that an active competition between the two branches of the fork is ongoing.

\end{itemize}

{\bf{Transition and Reward}}: After the execution of an action, the occurrence of each state transition is triggered by the creation of a new block (either by the adversary or by the honest network) and the corresponding transition probability is the probability of the block created by the adversary ($\alpha $) or by the honest network ($1-\alpha $). The initial state is $\left( {1,0,irrelevant} \right)$ with probability $\alpha $  or $\left( {0,1,irrelevant} \right)$ with probability $1-\alpha $. Different actions performed by the adversary will have different effects on the state transitions. The specific description is as follows:

\begin{itemize}
\item The state transitions after the execution of action $adopt$: By executing the $adopt$ action, the adversary accepts all the blocks on the branch mined by the honest network and mines on the latest block on the honest chain together with the honest network. An illustrating example of the state transitions after the execution of action $adopt$ is given in Fig. \ref{fig_adpot}. As shown in Fig. \ref{fig_adpot}, with the probability of $\alpha$, the adversary can successfully mine the next block and then the state transits to $\left( {1,0,relevant} \right)$; with the probability of $1-\alpha$, the honest network can successfully mine the next block and then the state transits to $\left( {0,1,relevant} \right)$.

\begin{figure}[h]
\centering
\includegraphics[width=3in]{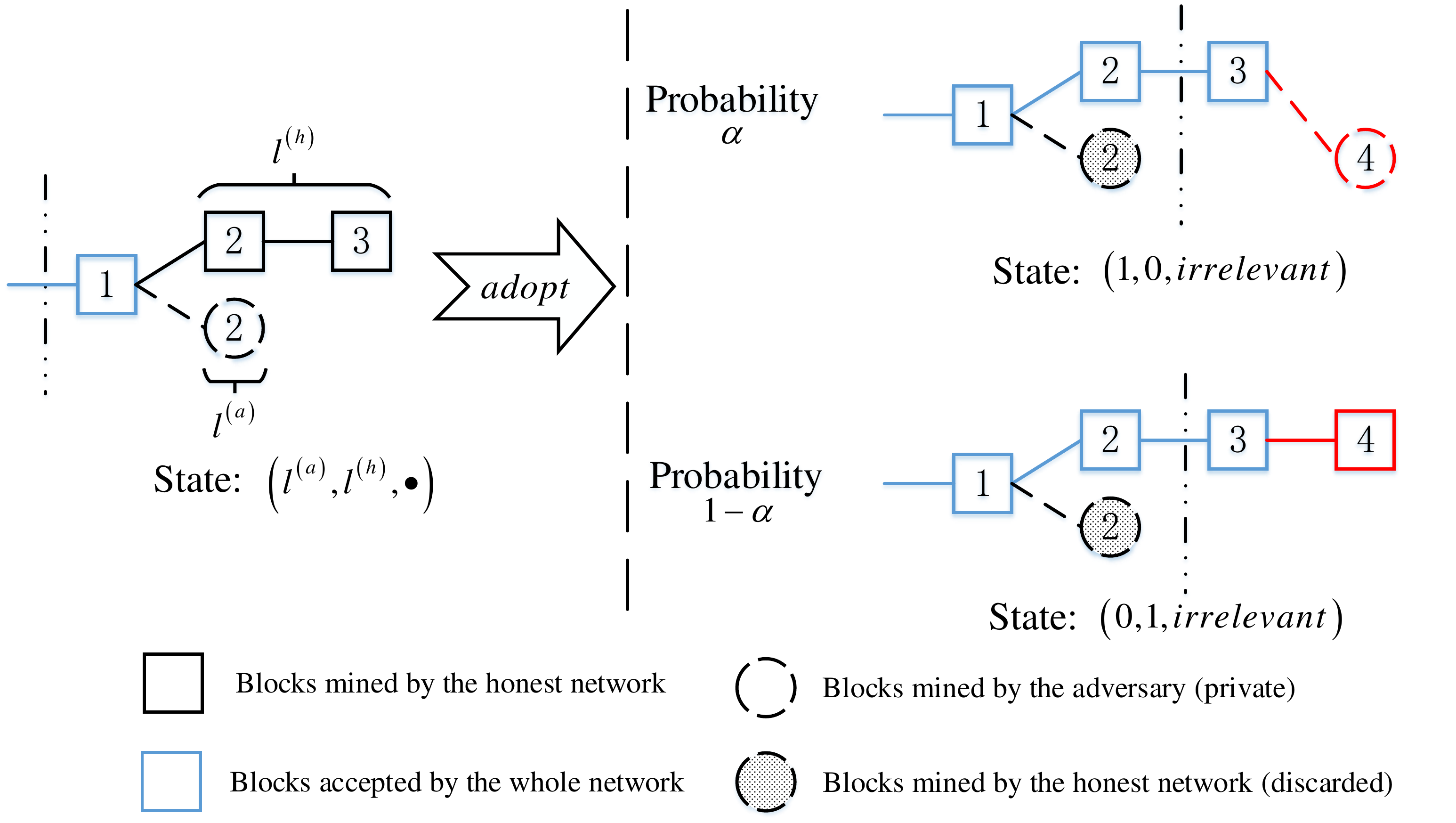}
\caption{An illustrating example of the state transitions after the execution of action $adopt$.} \label{fig_adpot}
\end{figure}

\item The state transition after the execution of action $override$: The adversary can only perform action $override$ when the number of the blocks on its private branch is greater than the number of the blocks on the honest branch (i.e., when ${l^{\left( a \right)}} > {l^{\left( h \right)}}$). By performing $override$, the adversary publishes ${l^{\left( h \right)}} + 1$
blocks from its private branch to overwrite the latest ${l^{\left( h \right)}}$ blocks on the honest branch. After that, the branch of the adversary becomes the main chain and the whole network mines on the latest block of the adversary's branch. An illustrating example of the state transitions after the execution of action $override$ is given in Fig. \ref{fig_override}. As shown in Fig. \ref{fig_override}, the adversary has the probability of $\alpha $ to successfully mine the next block and makes the state transit to $\left( {{l^{\left( a \right)}} - {l^{\left( h \right)}},0,irrelevant} \right)$; the honest network has the probability of $1-\alpha $ to successfully mine the next block and makes the state transit to $\left( {{l^{\left( a \right)}} - {l^{\left( h \right)}}-1,1, relevant} \right)$.
	
\begin{figure}[h]
\centering
\includegraphics[width=3in]{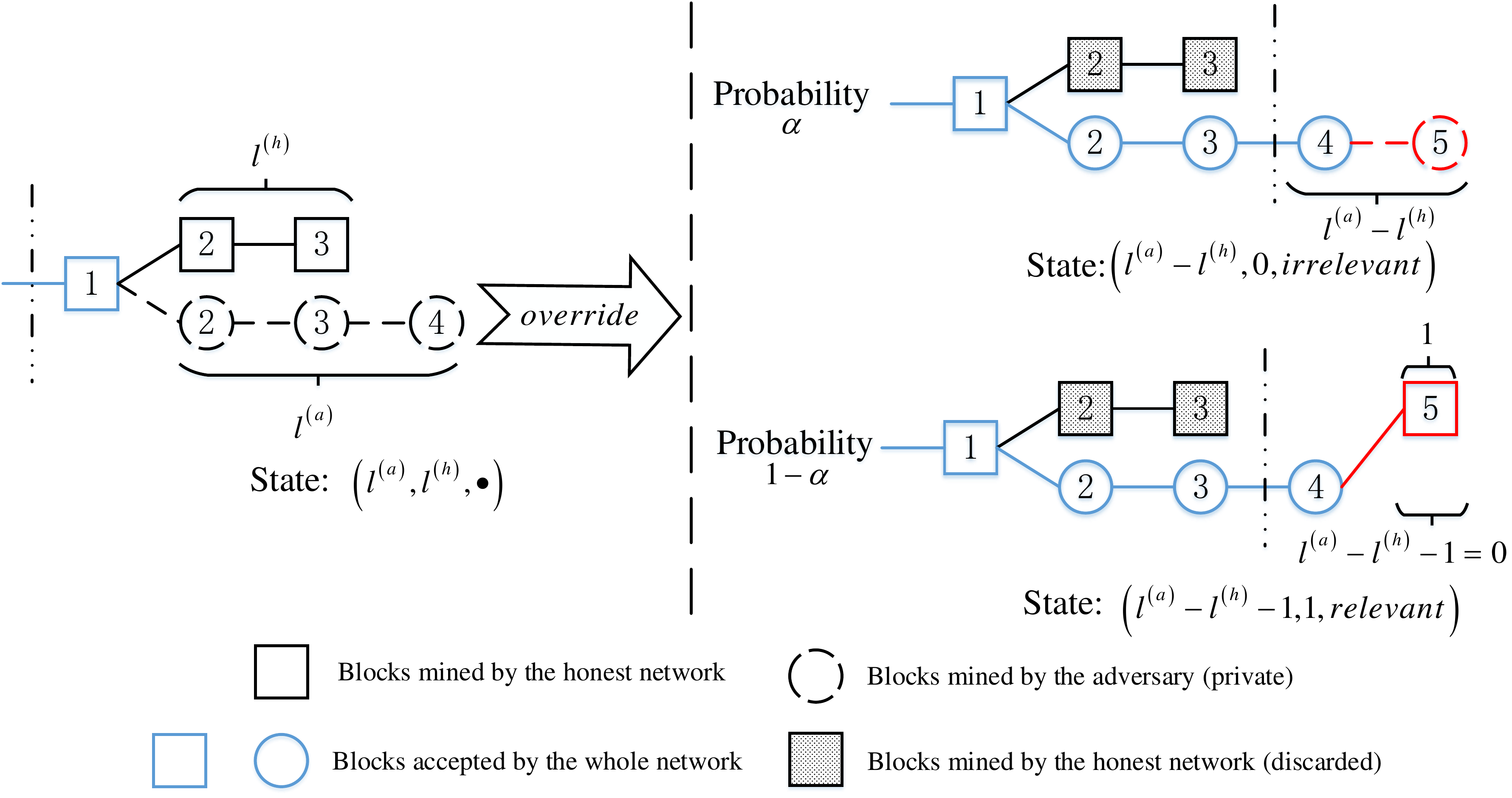}
\caption{An illustrating example of the state transitions after the execution of action $override$.} \label{fig_override}
\end{figure}

\item The state transition after the execution of action $match$: The $match$ action can only be executed when $fork = relevant$ and when the number of blocks on the private branch of the adversary is greater than or equal to the number of blocks on the public branch of the honest network (i.e., when ${l^{\left( a \right)}}  \ge  {l^{\left( h \right)}}$). After the adversary performs the $match$ action, a fork will be formed on the blockchain that is observed by all the miners. After that, the adversary is still mining on its own branch; however, due to the fork, a $\gamma $ fraction of the honest network will mine on the branch published by the adversary, and the other $1-\gamma $ fraction of the honest network will mine on the branch published by the honest network. An illustrating example of the state transitions after the execution of action $match$ is given in Fig. \ref{fig_match}. As shown in Fig. \ref{fig_match}, the next block may be published by the adversary on its own branch such that the state transits to $\left( {{l^{\left( a \right)}} + 1,{l^{\left( h \right)}},active} \right)$ with the probability of $\alpha$; the next block may be published by the honest network on the branch of the adversary such that the state transits to $\left( {{l^{\left( a \right)}} - {l^{\left( h \right)}},1,relevant} \right)$ with the probability of $\gamma \left( {1 - \alpha } \right)$; the next block may be published by the honest network on the branch of the honest network such that the state transits to $\left( {{l^{\left( a \right)}},{l^{\left( h \right)}} + 1,relevant} \right)$ with the probability of $\left( {1 - \gamma } \right)\left( {1 - \alpha } \right)$. We must emphasize that after the execution of action $match$, among the ${l^{\left( a \right)}}$ blocks of the adversary, some of the blocks may be private while other blocks are public. Which parts of blocks are private/public are implied by the state implicitly. For example, suppose that the previous state is $\left( {{l^{\left( a \right)}},{l^{\left( h \right)}},relevant} \right)$ with ${l^{\left( a \right)}} > {l^{\left( h \right)}}$ (as illustrated in the left part of Fig. \ref{fig_match}) and the action $match$ is performed. If the adversary subsequently mines a new block on its own branch, then the state changes to $\left( {{l^{\left( a \right)}} + 1,{l^{\left( h \right)}},active} \right)$, where there are ${l^{\left( a \right)}} + 1 - {l^{\left( h \right)}}$ private blocks and ${l^{\left( h \right)}}$ public blocks among the ${l^{\left( a \right)}} + 1$ blocks owned by the adversary (as illustrated by the first case in the right part of Fig. \ref{fig_match}). If the honest miners mine a new block on the adversary’s branch, the state changes to  $\left( {{l^{\left( a \right)}} - {l^{\left( h \right)}}, 1, relevant} \right)$, where there are ${l^{\left( a \right)}} - {l^{\left( h \right)}}$ private block left for the adversary (as illustrated by the second case in the right part of Fig. \ref{fig_match}). If the hones miners mine a new block on the honest network’s branch, the state changes to ${l^{\left( a \right)}} - {l^{\left( h \right)}}$, where there are ${l^{\left( a \right)}} - {l^{\left( h \right)}}$  private blocks and ${l^{\left( h \right)}}$ public blocks among the  ${l^{\left( a \right)}}$ blocks owned by the adversary (as illustrated by the third case in the right part of Fig. \ref{fig_match}).

\begin{figure}[h]
\centering
\includegraphics[width=3in]{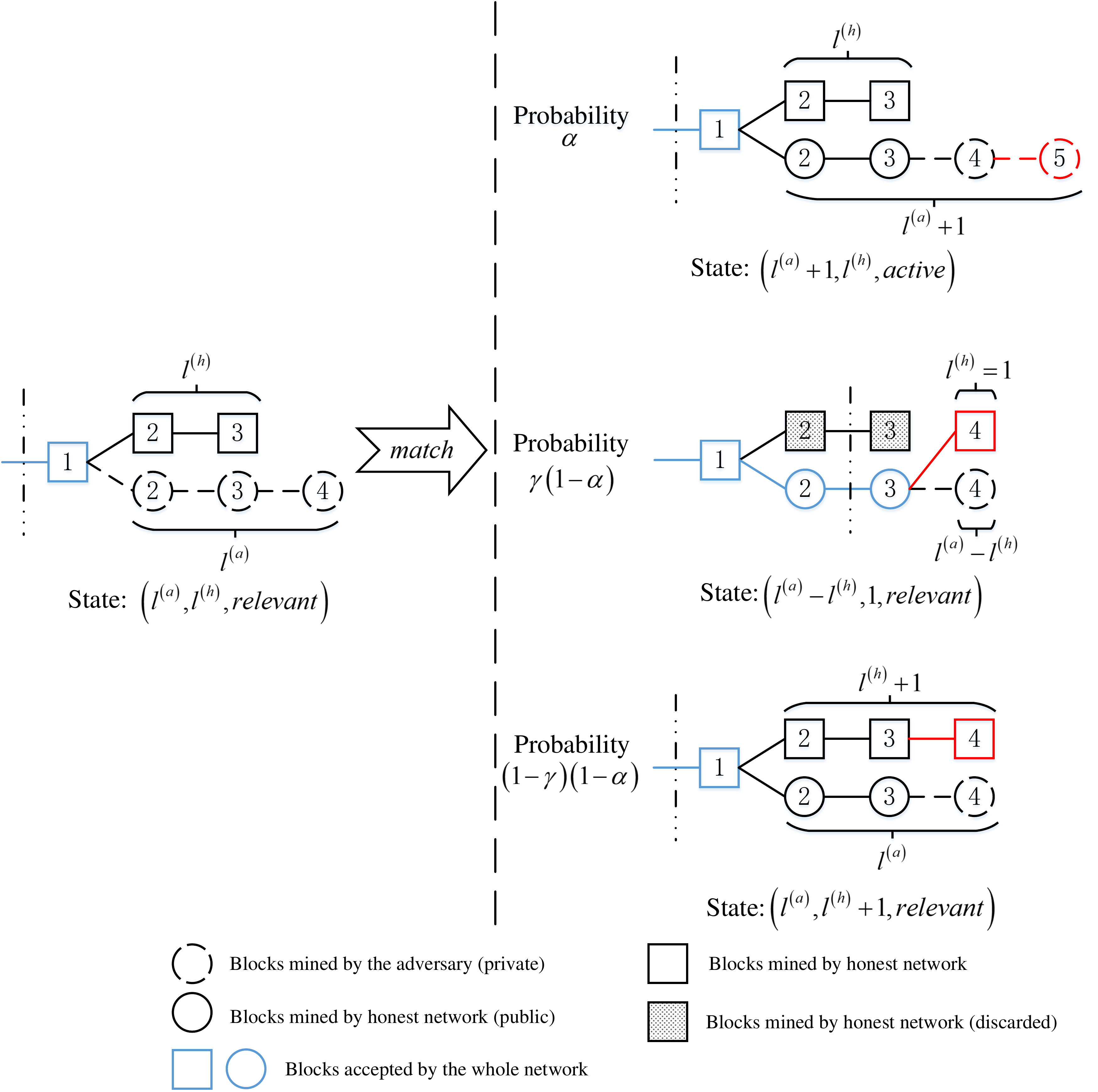}
\caption{An illustrating example of the state transitions after the execution of action $match$.} \label{fig_match}
\end{figure}

\item The state transition triggered by action $wait$: The $wait$ action means that the adversary does not perform any actions and continues to mine on its private branch. After the action $wait$ is executed, if $fork \ne active$, the adversary and the honest network mine on their own  branches respectively. An illustrating example of the state transitions after the execution of action $match$ when $fork \ne active$ is given in Fig. \ref{fig_wait}. As shown in Fig. \ref{fig_wait}, when $fork \ne active$, the next new block may be mined by the adversary on its own private branch such that the state changes to  $\left( {{l^{\left( a \right)}} + 1,{l^{\left( h \right)}},irrelevant} \right)$ with the probability of $\alpha$; or the next new block may be mined by the honest network on the public branch such that the state changes to $\left( {{l^{\left( a \right)}},{l^{\left( h \right)}} + 1,relevant} \right)$  with the probability of $1-\alpha$. After the action $wait$ is executed, if $fork = active$, due to the fork that can be observed by the whole network, the mining behaviors of all miners are the same as that after the execution of the $match$ action. An illustrating example of the state transitions after the execution of action $match$ when $fork = active$ is given in Fig. \ref{fig_wait2}. \color{black}{As shown in Fig. \ref{fig_wait2}, when $fork = active$, the next new block may be mined by the adversary on its own branch such that the state changes to  $\left( {{l^{\left( a \right)}} + 1,{l^{\left( h \right)}},active} \right)$ with probability $\alpha$; or the next new block may be mined by the honest network on the branch of the adversary such that the state changes to $\left( {{l^{\left( a \right)}} - {l^{\left( h \right)}},1,relevant} \right)$ with probability $\gamma \left( {1 - \alpha } \right)$;  or the next new block may be mined by the honest network on the branch of the adversary such that the state changes to $\left( {{l^{\left( a \right)}},{l^{\left( h \right)}} + 1,relevant} \right)$ with probability $\left( {1 - \gamma } \right)\left( {1 - \alpha } \right)$.}

\begin{figure}[h]
\centering
\includegraphics[width=3in]{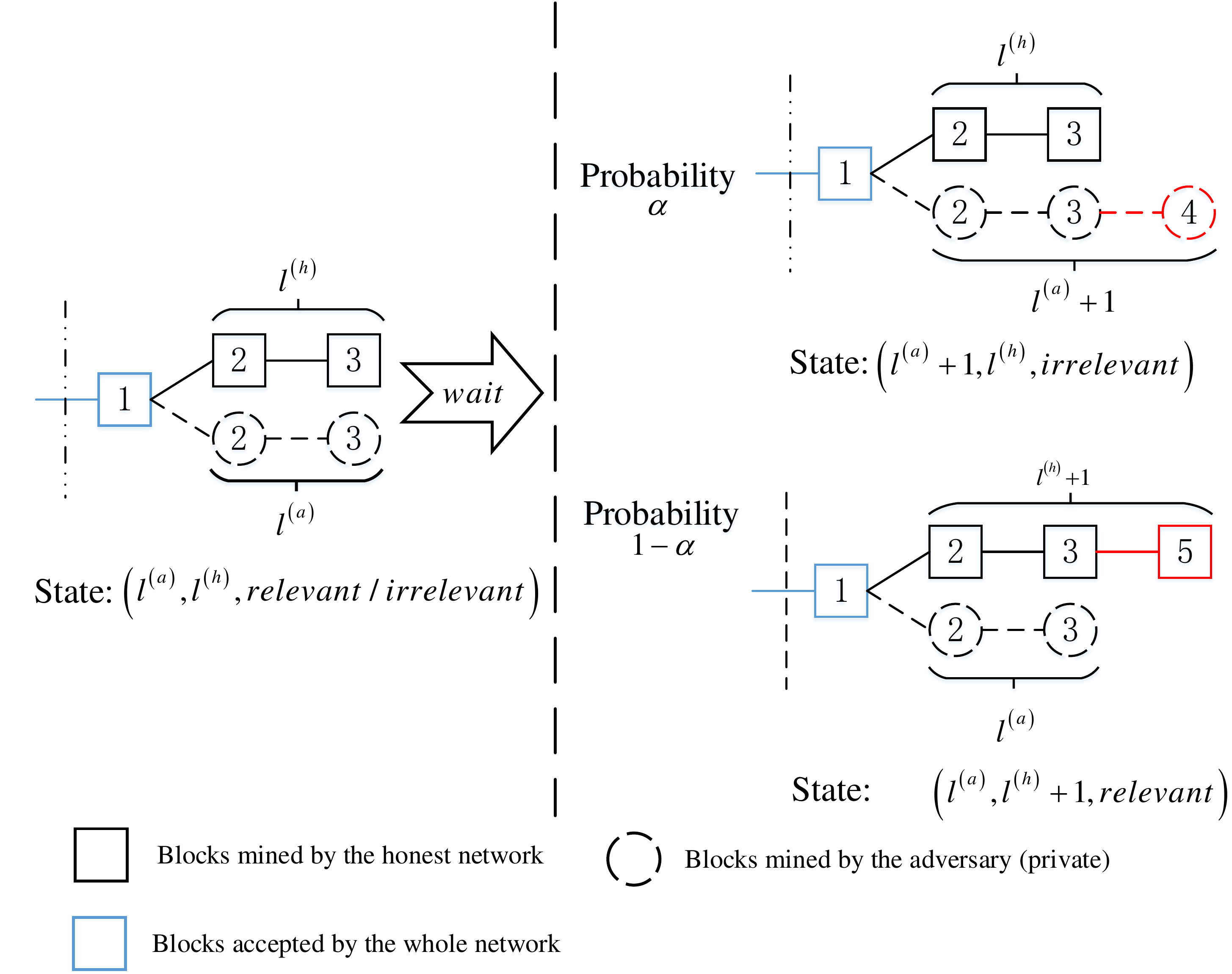}
\caption{An illustrating example of the state transitions after the execution of action $wait$ when $fork \ne active$.} \label{fig_wait}
\end{figure}

\begin{figure}[h]
	\centering
	\includegraphics[width=3in]{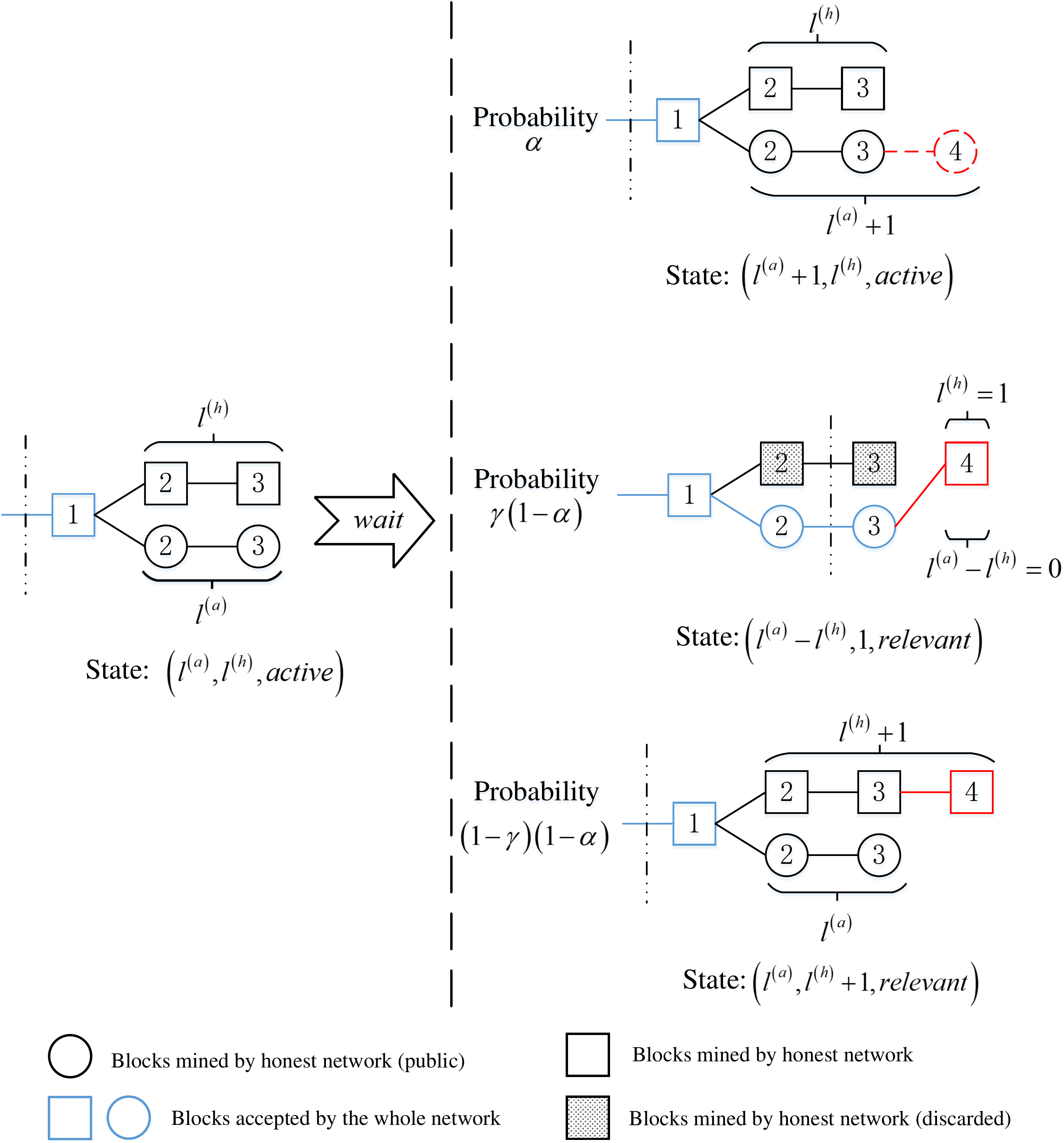}
	\caption{An illustrating example of the state transitions after the execution of action $wait$ when $fork = active$.} \label{fig_wait2}
\end{figure}

\end{itemize}

The reward is given as a tuple $\left( {{r^{\left( a \right)}},{r^{\left( h \right)}}} \right)$, where ${r^{\left( a \right)}}$ denotes the number of blocks mined by the adversary and accepted by the whole network, and ${r^{\left( h \right)}}$  denotes the number of blocks mined by the honest network and accepted by the whole network. The state transitions and reward matrices are given in TABLE I.

\begin{table*}
	\centering\centering
	\caption{The state transitions and reward matrices of the MDP mining model.}
	\label{table}
	\begin{tabular}{|l|l|l|l|}
		\hline
		Current State, Action  & Next State  &   Transition Probability  & Reward                  \\ \hline
		
		\multirow{2}{*}{$\left( {{l^{\left( a \right)}},{l^{\left( h \right)}}, \bullet } \right),adopt$} & $\left( {1,0,irrelevant} \right)$ & $\alpha$
		& \multirow{2}{*}{ $\left( {0,{l^{\left( h \right)}}} \right)$
		} \\ \cline{2-3}	
		&$\left( {0,1,irrelevant} \right)$  & $1-\alpha$  &       \\ \hline
		
		\multirow{2}{*}{$\left( {{l^{\left( a \right)}},{l^{\left( h \right)}}, \bullet } \right),override$} & $\left( {{l^{\left( a \right)}} - {l^{\left( h \right)}} ,0,irrelevant} \right)$ & $\alpha$
		& \multirow{2}{*}{ $\left( {{l^{\left( h \right)}} + 1,0} \right)$
		} \\ \cline{2-3}	
		&$\left( {{l^{\left( a \right)}} - {l^{\left( h \right)}} - 1,1,relevant} \right)$  & $1-\alpha$  &      \\ \hline

		$\left( {{l^{\left( a \right)}},{l^{\left( h \right)}},irrelevant} \right),wait$	&  $\left( {{l^{\left( a \right)}} + 1,{l^{\left( h \right)}},irrelevant} \right)$
		& $\alpha $  &  $\left( {0,0} \right)$  \\ \cline{2-4} 
		
		$\left( {{l^{\left( a \right)}},{l^{\left( h \right)}},relevant} \right),wait$ & $\left( {{l^{\left( a \right)}},{l^{\left( h \right)}} + 1,relevant} \right)$ & $1 - \alpha $ &   $\left( {0,0} \right)$
		\\ \hline

		$\left( {{l^{\left( a \right)}},{l^{\left( h \right)}},active} \right),wait$ & $\left( {{l^{\left( a \right)}} + 1,{l^{\left( h \right)}},active} \right)$ & $\alpha $ &   $\left( {0,0} \right)$                 \\ \cline{2-4} 
		
		$\left( {{l^{\left( a \right)}},{l^{\left( h \right)}},relevant} \right),match$ & $\left( {{l^{\left( a \right)}} - {l^{\left( h \right)}},1,relevant} \right)$ & $\gamma \left( {1 - \alpha } \right)$ &    $\left( {{l^{\left( h \right)}},0} \right)$  \\ \cline{2-4}

		& $\left( {{l^{\left( a \right)}},{l^{\left( h \right)}} + 1,relevant} \right)$ & $\left( {1 - \gamma } \right)\left( {1 - \alpha } \right)$  &        $\left( {0,0} \right)$              \\ \hline
	\end{tabular}
	\\
	\footnotesize{The action $override$ is allowed when ${l^{\left( a \right)}} > {l^{\left( h \right)}}$; the action $match$ is allowed when ${l^{\left( a \right)}} \ge {l^{\left( h \right)}}$.}
	\label{tab1}
\end{table*}


{\bf{Objective Function}}: The objective of the adversary is to find the optimal mining strategy that can earn as much reward as possible. Since blockchain keeps adjusting the mining difficulty (i.e., the mining target on the RHS of inequality (1)) to ensure that on average one valid block is introduced to the overall blockchain per \emph{valid block interval} (e.g., one block per 10 minutes for Bitcoin, and per 10-20 seconds for Ethereum), the mining objective of the adversary is not to maximize its absolute cumulative reward, but to maximize the ratio of its cumulative rewards over the cumulative rewards of the whole network (i.e., the cumulative rewards of the whole network advance by one reward per block interval---rewards of all miners/Time is fixed to 1 per block interval; then maximizing adversary rewards/Time is equivalent to maximizing the ratio of adversary rewards/Time to rewards of all miners/Time = adversary rewards/rewards of all miners). We emphasize that blocks mined by the adversary and the honest network that are discarded due to losing out in the competition are not considered as having been successfully introduced to the blockchain. Thus, the principle behind the strategy of the adversary is to maximize the number of blocks mined by the honest network that are later discarded while reducing its own discarded blocks.

As in \cite{sapirshtein2016optimal}, we define the following $relative$ $mining$ $gain$ ($RMG$) as the objective function for blockchain mining:
\begin{equation}
RMG = E\left[ {\mathop {\lim }\limits_{T \to \infty } {{\sum\nolimits_{\tau  = t}^{t + T - 1} {r_{\tau  + 1}^{\left( a \right)}} } \over {\sum\nolimits_{\tau  = t}^{t + T - 1} {r_{\tau  + 1}^{\left( a \right)}}  + \sum\nolimits_{\tau  = t}^{t + T - 1} {r_{\tau  + 1}^{\left( h \right)}} }}} \right]
\end{equation}
where $\left( {r_t^{\left( a \right)},r_t^{\left( h \right)}} \right)$ is the tuple of rewards issued in the block interval $t$, $T$  is the size of the observing window. The objective of the adversary is to maximize this relative mining gain. 

Under the above MDP mining model, we can now interpret honest mining, selfish mining \cite{eyal2018majority}, lead stubborn mining \cite{nayak2016stubborn} as examples of different mining strategies. 

{\bf{Honest Mining}}: For honest mining, miners will follow the rule of the longest chain extension. Thus, they will not maintain a private chain: when they have a new block, they will immediately publish it. The honest mining strategy can be written as 
\begin{equation}
HM\left( {{l^{\left( a \right)}},{l^{\left( h \right)}}, \bullet } \right) = \left\{ {\begin{array}{*{20}{c}}
	\begin{array}{l}
	adopt \\ 
	wait \\ 
	\end{array} & \begin{array}{l}
	{l^{\left( a \right)}} < {l^{\left( h \right)}} \\ 
	{l^{\left( a \right)}} = {l^{\left( h \right)}} \\ 
	\end{array}  \\
	{override} & {{l^{\left( a \right)}} > {l^{\left( h \right)}}}  \\
	\end{array}} \right.
\end{equation}
where we note that ${l^{\left( a \right)}}$,  ${l^{\left( h \right)}}$ can only take a value of 0 or 1. 

{\bf{Selfish Mining}}: The main idea of selfish mining \cite{eyal2018majority} is described as follows. If one block is found by the adversary, it does not publish it immediately and it keeps mining on its private chain. When the adversary already has one private block and then honest network publishes one block (immediately after an honest miner mines a new block), the adversary chooses to publish its block to match the honest network. This causes $\gamma (1 - \alpha )$  computing power of the honest network to mine on the adversary’s chain. When the adversary already has some private blocks and then honest network catches up with only one block less than the adversary (${l^{\left( h \right)}} = {l^{\left( a \right)}} - 1 \ge  1$),  the adversary overrides the honest network's block by publishing all its blocks. The selfish mining strategy can be written as 
\begin{equation}
SM\left( {{l^{\left( a \right)}},{l^{\left( h \right)}}, \bullet } \right) = \left\{ {\begin{array}{*{20}{c}}
	\begin{array}{l}
	adopt \\ 
	match \\ 
	override \\ 
	\end{array} & \begin{array}{l}
	{l^{\left( a \right)}} < {l^{\left( h \right)}} \\ 
	{l^{\left( a \right)}} = {l^{\left( h \right)}} = 1 \\ 
	{l^{\left( h \right)}} = {l^{\left( a \right)}} - 1 \ge 1 \\ 
	\end{array}  \\
	{wait} & {otherwise}  \\
	\end{array}} \right.
\end{equation}

{\bf{ Lead Stubborn Mining}}: Lead stubborn mining \cite{nayak2016stubborn} is different from selfish mining in the following way. A lead stubborn miner always publishes one block from its private chain to match with the honest network when the honest network mines a new block if ${l^{\left( a \right)}} \ge {l^{\left( h \right)}}$. The adversary never executes the action  override. The lead stubborn mining can be written as
\begin{equation}
\begin{array}{l}
LSM\left( {{l^{\left( a \right)}},{l^{\left( h \right)}},fork} \right) \\ 
= \left\{ {\begin{array}{*{20}{c}}
	\begin{array}{l}
	adopt \\ 
	match \\ 
	\end{array} & \begin{array}{l}
	{l^{\left( a \right)}} < {l^{\left( h \right)}},\forall fork \\ 
	otherwise \\ 
	\end{array}  \\
	{wait} & {{l^{\left( a \right)}} > {l^{\left( h \right)}},fork = irrelevant}  \\
	\end{array}} \right. \\ 
\end{array}
\end{equation}
It is shown that this lead stubborn mining can achieve higher profits than selfish mining \cite{nayak2016stubborn}.

{\bf{Optimal Mining}}: Although there are many possible mining strategies that can obtain profits higher than honest mining, the optimal mining strategy is not obvious. Since the state-action space of the MDP is huge, it is not straightforward to derive the optimal mining strategy. The relative mining gain objective (2) is a nonlinear function of the rewards, and thus the corresponding MDP cannot be solved using standard MDP solvers to give the optimal mining strategy. To solve this problem, \cite{sapirshtein2016optimal} first transformed the MDP with the nonlinear objective to a family of MDPs with linear objectives, and then employed a standard MDP solver combined with a numerical search over the family of MDPs to find the optimal mining strategy. As shown in \cite{sapirshtein2016optimal}, its solution indeed can find the optimal mining strategy.  However, the solution of \cite{sapirshtein2016optimal} is model-based approach: it must know the parameters that characterize the MDP model exactly (i.e., the computing power distribution $\alpha $, the communication capability $\gamma $). In real blockchain networks, these parameters are not easy to obtain and may change over time, hindering the use of the solution proposed in \cite{sapirshtein2016optimal}. We propose a model-free approach that solves the MDP with the nonlinear objective using RL.

\section{Mining Through RL}

This section first provides preliminaries for RL and then presents a new RL algorithm that can derive the optimal mining strategy without knowing the parameters of the environment. We propose the new RL mining algorithm based on Q-learning, one popular algorithm from the RL family.

\subsection{Preliminaries for Original Reinforcement Learning Algorithm}
\begin{figure}[t]
	\centering
	\includegraphics[width=3in]{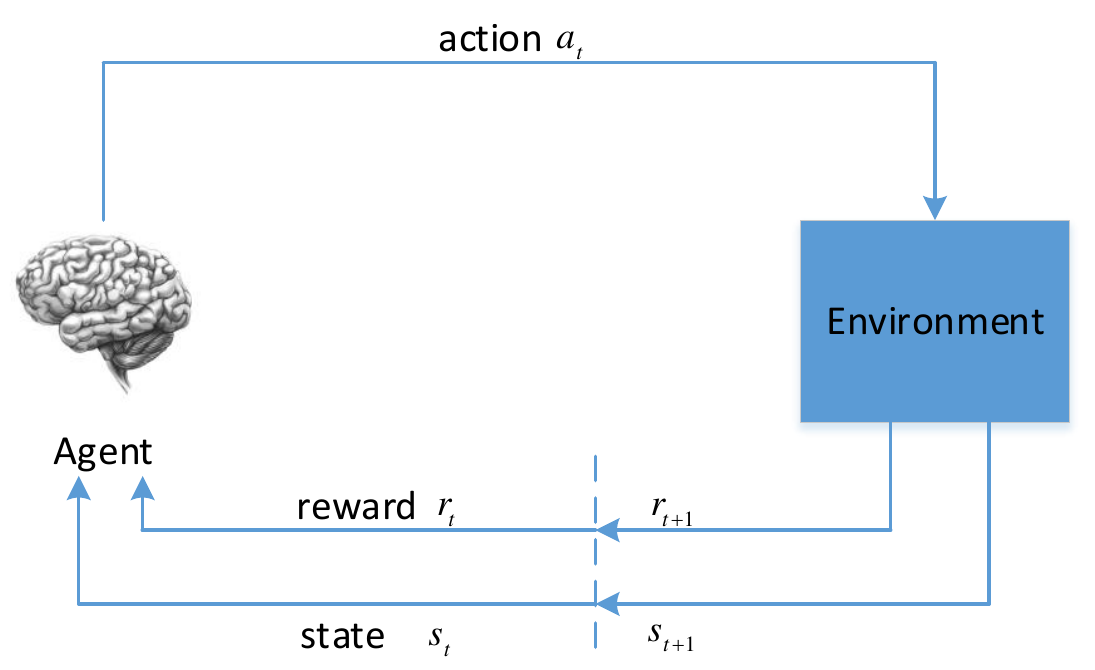}
	\caption{The agent-environment interaction process of RL algorithm.} \label{rl_alg}
\end{figure}

In RL, an agent interacts with an environment in a sequence of discrete time steps, $t = 0,1,2,...,$ as shown in Fig. \ref{rl_alg}. At time $t$, the agent observes the state of the environment, ${s_t}$; it then takes an action, ${a_t}$. As a result of the state-action pair, $({s_t},{a_t})$, the agent receives a scalar reward $r_{t+1}$, and the environment moves to a new state $s_{t+1}$ at time $t+1$. Based on $s_{t+1}$, the agent then decides the next action $a_{t+1}$. The goal of the agent is to effect a series of rewards $\{ {r_t}\} _{t = 1,2,...}^{}$ through its actions to maximize some performance criterion. For example, for Q-learning \cite{watkins1992q}, the performance criterion to be maximized at time $t$ is the discounted accumulated rewards going forward ${R_t} = \sum\nolimits_{\tau  = t}^\infty  {{\lambda ^{\tau  - t}}{r_{\tau {\rm{ + }}1}}} $, where $\lambda  \in \left( {0,1} \right)$ is a discount factor for weighting future rewards \cite{sutton2018reinforcement}. In general, the agent takes actions according to some decision policy $\pi $. RL methods specify how the agent changes its policy as a result of its experiences. With sufficient experiences, the agent can learn an optimal decision policy ${\pi ^ * }$ to maximize the long-term accumulated reward \cite{sutton2018reinforcement}. 

The desirability of state-action pair $({s_t},{a_t})$ under a decision policy decision ${\pi }$ is captured by a Q function, defined as  $Q\left( {s,a} \right) = \left[ {{R_t}\left| {{s_t} = s,{a_t} = a,\pi } \right.} \right]$, i.e., the expected discounted accumulated reward going forward given the current state-action pair $({s_t},{a_t})$. The optimal decision policy  ${\pi ^ * }$  is  one that maximizes Q function. In Q-learning, the goal of the agent is to learn the optimal policy  ${\pi ^ * }$  through an online-iterative process by observing the rewards while it takes action in successive time steps. In particular, the agent maintains the Q function, $Q(s,a)$, for all state-action pairs $\left( {s,a} \right)$, in a tabular form. 

Let $q\left( {s,a} \right)$ be the estimated action-value function during the iterative process.  At time step $t$,  given state ${s_t}$, the agent selects a greedy action ${a_t} = \arg {\max _a}q({s_t},a)$
based on its current Q function. This will cause the system to return a reward ${r_{t + 1}}$ and move to state  ${s_{t + 1}}$. The experience at time step $t$ is captured by the quadruplet ${e_t} = ({s_t},{a_t},{r_{t + 1}},{s_{t + 1}})$. At the end of time step $t$, experience ${e_t}$ is used to update $q({s_t},{a_t})$ for entry $\left( {{s_t},{a_t}} \right)$ as follows:
\begin{equation}
q\left( {{s_t},{a_t}} \right) \leftarrow \left( {1 - \beta } \right)q\left( {{s_t},{a_t}} \right) + \beta \left[ {{r_{t + 1}} + \lambda \mathop {\max }\limits_{a'} q\left( {{s_{t + 1}},a'} \right)} \right]
\end{equation}
where $\beta  \in (0,1]$ is a parameter that governs the learning rate. Q-learning learns from experiences gathered over time, ${\{ {e_t}\} _{t = 0,1,...}}$, through the iterative process in (6).  Note that Q-learning is a model-free learning framework in that it tries to learn the optimal policy without having a model that describes the operating behavior of the environment beyond what can be observed through the experiences. 

As a deviation from the above description, a caveat in Q-learning is that the so-called  $\varepsilon$-greedy algorithm is often adopted in action selection. For the $\varepsilon$-greedy algorithm, the action ${a_t} = \arg {\max _a}q({s_t},a)$  is only chosen with probability  $1 - \varepsilon $. With probability  $\varepsilon $, a random action is chosen uniformly from the set of possible actions.  This is to avoid the algorithm from zooming in to a local optimal policy and to allow the agent to explore a wider spectrum of different actions in search of the optimal policy \cite{sutton2018reinforcement}.

It has been shown that in a stationary environment that can be fully captured by an MDP, the Q-values will converge to optimality if the learning rate decays appropriately and each action in the state-action pair  $\left( {s,a} \right)$ is executed an infinite number of times in the process \cite{sutton2018reinforcement}.

\subsection{New Reinforcement Learning Algorithm for Mining}

The original RL algorithm as presented in Section IV.A cannot be directly applied to maximize the mining objective function expressed in (2); there is one fundamental obstacle that must be overcome. The obstacle is the nonlinear combination of the rewards in the objective function. The original RL algorithm can only maximize an objective that is a linear function of the scalar rewards, e.g., the weighted sum of scalar rewards. To address  this issue, we put forth a new algorithm that aims to optimize the original mining objective: the  multi-dimensional RL algorithm.


We formulate the multi-dimensional RL algorithm as follows. At \emph{mined block interval} \footnote{A mined block interval is different from a valid block interval. A valid block interval separates two valid blocks that are ultimately adopted by the blockchain. The average duration of a valid block interval is a constant in many blockchain systems (e.g., 10 min in bitcoin).  The average duration of the valid block interval is defined by the system designer and its constancy is maintained by adjusting the mining target. A mined block interval separated two mined (by either the adversary of the honest network), regardless of whether the blocks becomes valid later. In the MDP model, each transition is triggered by the mining of a new block. Thus the average duration of a mined block interval is the average time separates two adjacent transitions.  Due to the actions of the adversary, some of the mined blocks (by the adversary of the honest network) may be discarded later. } $t$ ($t = 0,1,2, \cdots $), the state ${s_t} \in S$ takes a value from the state space $S$ as defined in the MDP model of blockchain mining, and the action ${a_t} \in A$ is chosen from the action space $A$. The state transition occurs according to TABLE I. The reward is the pair $\left( {r_{t + 1}^{\left( a \right)},r_{t + 1}^{\left( h \right)}} \right)$ whose value is assigned according to TABLE I. The experience at the end of mined block interval $t$ is given by ${e_t} = \left( {{s_t},{a_t},{s_{t + 1}},r_{t + 1}^{\left( a \right)},r_{t + 1}^{\left( h \right)}} \right)$. The objective of the multi-dimensional RL algorithm is to maximize the relative mining gain as expressed in (2).  

For a state-action pair $({s_t},{a_t})$, instead of maintaining an action-value scalar $Q(s,a)$, the multi-dimensional RL algorithm maintains an action-value pair $\left( {{Q^{(a)}}(s,a),{Q^{(h)}}(s,a)} \right)$ corresponding to the Q function values of the adversary and the honest network, respectively. The Q functions defined by Q learning are the expected cumulative discounted rewards. Specifically,  ${Q^{(a)}}(s,a)$  and ${Q^{(h)}}(s,a)$ are defined as   
\begin{equation}
\begin{gathered}
{Q^{\left( a \right)}}(s,a) = E\left[ {\mathop {\lim }\limits_{T \to \infty } \sum\nolimits_{\tau  = t}^T {{\lambda ^{\tau  - t}}r_{\tau  + 1}^{\left( a \right)}} \left| {{s_t} = s,{a_t} = a,\pi } \right.} \right] \hfill \\
{Q^{\left( h \right)}}(s,a) = E\left[ {\mathop {\lim }\limits_{T \to \infty } \sum\nolimits_{\tau  = t}^T {{\lambda ^{\tau  - t}}r_{\tau  + 1}^{\left( h \right)}{\kern 1pt} {\kern 1pt} {\kern 1pt} \left| {{\kern 1pt} {\kern 1pt} {s_t} = s,{a_t} = a,\pi } \right.} } \right] \hfill \\ 
\end{gathered} 
\end{equation}
Suppose that at mined block interval $t$, the Q functions in (7) are estimated to be ${q^{\left( a \right)}}(s,a)$, ${q^{\left( h \right)}}(s,a)$. For action selection, we still adopt the  $\varepsilon $-greedy approach. To select the greedy action, we construct the following objective function:
\begin{equation}
f\left( {s,a} \right) = \frac{{{q^{\left( a \right)}}(s,a)}}
{{{q^{\left( a \right)}}(s,a) + {q^{\left( h \right)}}(s,a)}}
\end{equation}
After taking action $a_t$, the state transitions to  ${s_{t + 1}}$ and the reward pair $\left( {r_{t + 1}^{\left( a \right)},r_{t + 1}^{\left( h \right)}} \right)$ is issued. With the experience ${e_t} = \left( {{s_t},{a_t},{s_{t + 1}},r_{t + 1}^{\left( a \right)},r_{t + 1}^{\left( h \right)}} \right)$, the multi-dimensional RL algorithm updates the two Q functions as follows:
\begin{equation}
\begin{gathered}
{q^{\left( a \right)}}({s_t},{a_t}) \hfill \\
\leftarrow \left( {1 - \beta } \right){q^{\left( a \right)}}({s_t},{a_t}) + \beta \left[ {r_{t + 1}^{\left( a \right)} + \lambda {q^{\left( a \right)}}\left( {{s_{t + 1}},a'} \right)} \right] \hfill \\
{q^{\left( h \right)}}({s_t},{a_t}) \hfill \\
\leftarrow \left( {1 - \beta } \right){q^{\left( h \right)}}({s_t},{a_t}) + \beta \left[ {r_{t + 1}^{\left( h \right)} + \lambda {q^{\left( h \right)}}\left( {{s_{t + 1}},a'} \right)} \right] \hfill \\ 
\end{gathered} 
\end{equation}
where  $a' = \arg {\max _a}f\left( {{s_{t + 1}},a} \right)$. Note that the update rule of (9) is very similar to the update rule of Q learning, except that the greedy action $a'$ is chosen by maximizing the constructed objective function in (8) rather than maximizing the Q function itself as in Q learning. From the expressions in (7) and (8), we can verify that the adopted objective function in (8) is consistent with the relative mining gain objective function defined in (2), except the discount terms ${\lambda ^{\tau  - t}}$  used  in the computation of  the Q functions. The use of discount terms can ensure that the Q functions can converge to some bounded values; however, adding discount terms to the rewards will change the original mining objective. One simple way to ensure strict objective consistency is to set  $\lambda  = 1$. Although the setting of $\lambda  = 1$  will result in unbound values for the Q functions as the RL iteration gradually progresses to infinite time steps, this is not a big problem as long as the Q function values do not overflow during the execution of the algorithm. In practice, we can also set  $\lambda $ to be very close to one. 

The RL algorithm expressed by the Q function updates in (9) is our multi-dimensional RL algorithm.  We introduce one additional technical element to the  $\varepsilon $-greedy action selection, as explained in the next paragraph.

As described above, when we select the action, we adopt the $\varepsilon $-greedy strategy that allows us to select the current best action (${a_t} = \arg {\max _a}f\left( {{s_t},a} \right)$) with probability $1 - \varepsilon $ and to randomly select an action with probability $\varepsilon $. This random action selection is used to explore some unseen states and can avoid trapping at local optimal maximums. However, the tuning of parameter $\varepsilon $ is not straightforward. A large $\varepsilon $  reduces the possibility of trapping at local optimal maximums but it also decrease the average reward, since it wastes a fraction of the time to explore non-optimal states.  In our algorithm, we adopt the following strategy for dynamically tuning the parameter $\varepsilon $. Denote the number of times state  was visited by $V\left( {{s_t}} \right)$. Then, the  $\varepsilon $ parameter used at state $s_t$ for performing  $\varepsilon $-greedy action selection is given by 
\begin{equation}
\varepsilon \left( {{s_t}} \right) = \exp \left( { - \frac{{V\left( {{s_t}} \right)}}
	{{{T_\varepsilon }}}} \right)
\end{equation}
where ${T_\varepsilon }$ is a temperature parameter that governs how fast we gradually reduce the  $\varepsilon $ parameter. The pseudo-code of our multi-dimensional RL algorithm for blockchain mining is given in Algorithm 1.

\begin{algorithm}[!h]
	\caption{Multi-dimensional RL Algorithm for Blockchain Mining}\label{alg}
	\begin{algorithmic}
		\State Initialize ${q^{\left( a \right)}}\left( {s,a} \right) = 0,\forall s,\forall a$;
		\State Initialize ${q^{\left( h \right)}}\left( {s,a} \right) = 0,\forall s,\forall a$;
		\State Initialize $V\left( s \right) = 0,\forall s$;
		\State Initialize  ${T_\varepsilon }$, $\lambda$, $\beta$;
		
		\For{$ t=0,1,2,\cdots $}
		\State Receive $s_t$, $r_t$ from the blockchain environment;
		\State Generate action ${a_t}=$ \Call{SelectAction}{$s_t$};
		\State Input ${a_t}$ to the blockchain environment;
		\State Observe $s_{t+1}$, $r_{t + 1}^{\left( a \right)}$, $r_{t + 1}^{\left( h \right)}$
		from the blockchain environment;
		\State Compute $a' = \arg \mathop {\max }\limits_a \frac{{{q^{\left( a \right)}}\left( {{s_{t + 1}},a} \right)}}{{{q^{\left( a \right)}}\left( {{s_{t + 1}},a} \right) + {q^{\left( h \right)}}\left( {{s_{t + 1}},a} \right)}}$
		\State Update $$
		\begin{gathered}
		{q^{\left( a \right)}}({s_t},{a_t}) \hfill \\
		\leftarrow \left( {1 - \beta } \right){q^{\left( a \right)}}({s_t},{a_t}) + \beta \left[ {r_{t + 1}^{\left( a \right)} + \lambda {q^{\left( a \right)}}\left( {{s_{t + 1}},a'} \right)} \right] \hfill; \\
		{q^{\left( h \right)}}({s_t},{a_t}) \hfill \\
		\leftarrow \left( {1 - \beta } \right){q^{\left( h \right)}}({s_t},{a_t}) + \beta \left[ {r_{t + 1}^{\left( h \right)} + \lambda {q^{\left( h \right)}}\left( {{s_{t + 1}},a'} \right)} \right] \hfill ;\\ 
		\end{gathered} 
		$$			
		\EndFor

		\Procedure{SelectAction}{$s_t$}
	    		\State Compute $\varepsilon \left( {{s_t}} \right) = \exp \left( { - \frac{{V\left( {{s_t}} \right)}}{{{T_\varepsilon }}}} \right)$;
	    
\If{random $<\varepsilon \left( {{s_t}} \right)$}
   \State randomly select an action $a_t$ from $A$;
\Else
   \State ${a_t} = \arg \mathop {\max }\limits_{a \in A} \frac{{{q^{\left( a \right)}}\left( {{s_t},a} \right)}}{{{q^{\left( a \right)}}\left( {{s_t},a} \right) + {q^{\left( h \right)}}\left( {{s_t},a} \right)}}$;
\EndIf 
    \Return ${a_t}$
	\EndProcedure
	
	\end{algorithmic}
\end{algorithm}

\section{Performance Evaluations}

We have conducted simulations to investigate our proposed RL mining strategy. Following the simulation approach used in \cite{eyal2018majority}, we constructed a Bitcoin-like simulator that captures all the relevant PoW  network  details described in previous sections, except that the crypto puzzle solving processing was replaced by a Monte Carlo simulator that simulates the time required for block discovery without actually attempting to compute a hash function. We simulated 1000 miners mining at identical rates (i.e., they each  can have one simulated hash test at each time step of the Monte Carlo simulation). A subset of the 1000 miners ($1000\alpha $ miners) forms an adversary pool running a malicious mining strategy that co-exists with honest mining adopted by the other $1000(1-\alpha) $ miners. When co-existing with honest mining, the malicious mining strategy is one of the following mining strategies: i) our RL mining strategy, ii) the optimal mining strategy derived in \cite{sapirshtein2016optimal} or iii) the selfish mining strategy derived in \cite{eyal2018majority}. Upon encountering  two subchains of the same length, we divide the honest miners such that a fraction  $\gamma $ of  them mine on the attacking pool's branch while the rest mine on the other branch. The performance metric used is the relative mining gain (RMG) computed over a window consisting of ${T_w} = {10^5}$ time steps:   $\sum\nolimits_{\tau  = t}^{t + {T_w} - 1} {{{r_{\tau  + 1}^{\left( a \right)}} \mathord{\left/
			{\vphantom {{r_{\tau  + 1}^{\left( a \right)}} {\left( {\sum\nolimits_{\tau  = t}^{t + {T_w} - 1} {r_{\tau  + 1}^{\left( a \right)}}  + \sum\nolimits_{\tau  = t}^{t + {T_w} - 1} {r_{\tau  + 1}^{\left( h \right)}} } \right)}}} \right.
			\kern-\nulldelimiterspace} {\left( {\sum\nolimits_{\tau  = t}^{t + {T_w} - 1} {r_{\tau  + 1}^{\left( a \right)}}  + \sum\nolimits_{\tau  = t}^{t + {T_w} - 1} {r_{\tau  + 1}^{\left( h \right)}} } \right)}}} $. The hyper-parameters used in the RL algorithm are set to as $\lambda  = 0.999$, $\beta  = 0.05$.

\begin{figure}[t]
	\centering
	\includegraphics[width=3.2in]{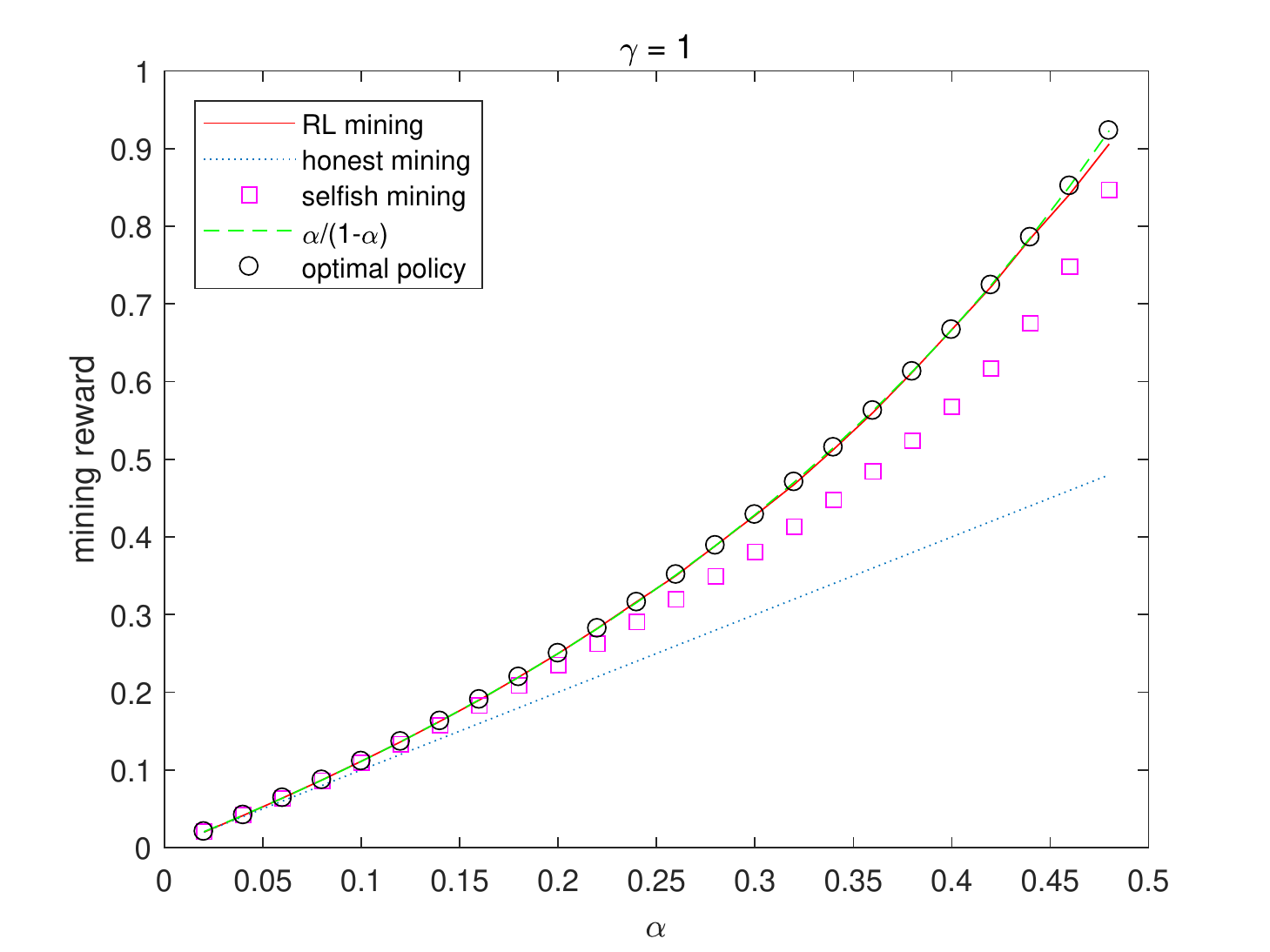}
	\caption{Achieved mining gain versus $\alpha $ for $\gamma =1$.} \label{fig_simu1}
\end{figure}

\begin{figure}[t]
	\centering
	\includegraphics[width=3.2in]{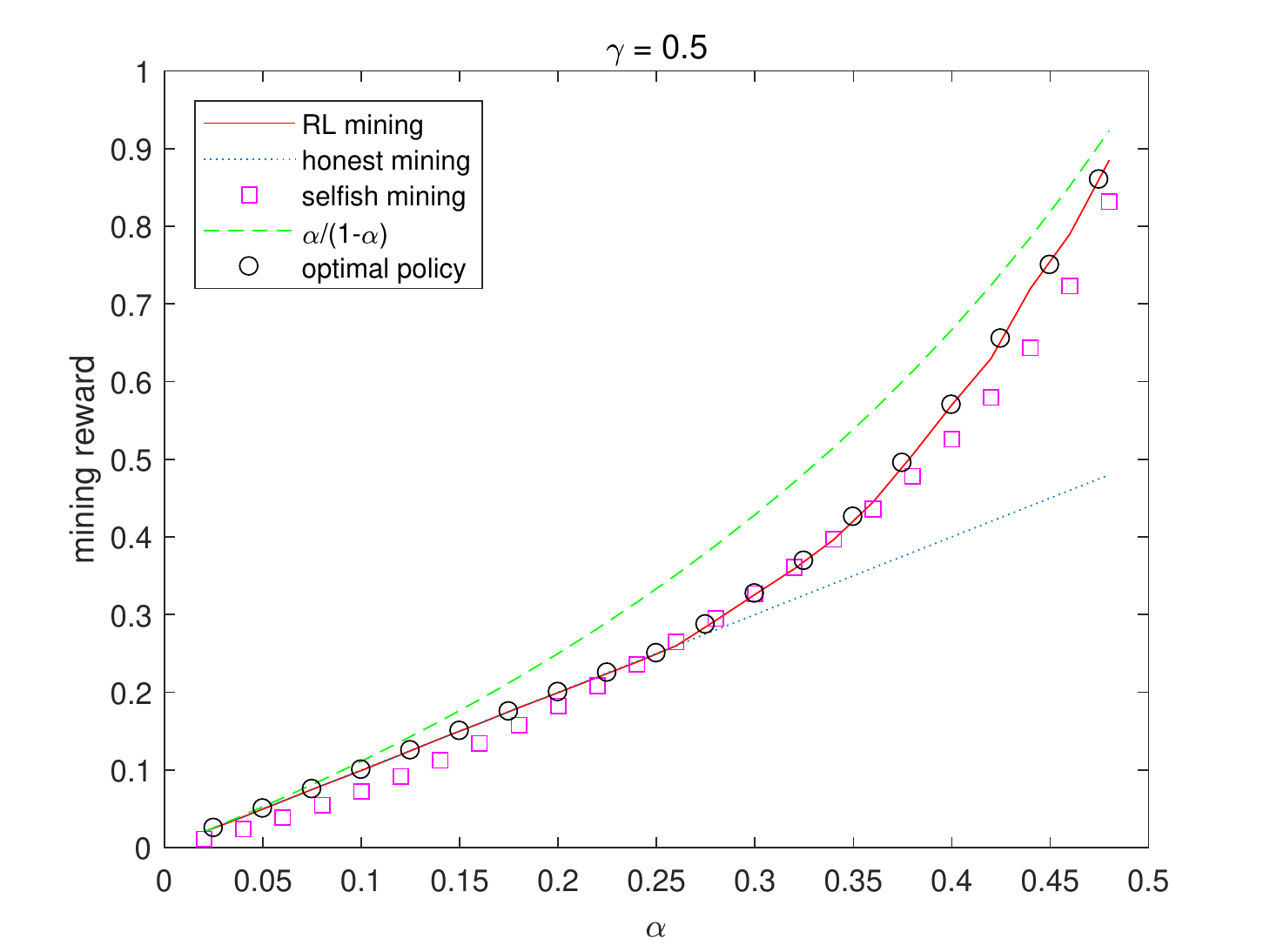}
	\caption{Achieved mining gain versus $\alpha $ for $\gamma =0.5$. } \label{fig_simu2}
\end{figure}

\begin{figure}[t]
	\centering
	\includegraphics[width=3.2in]{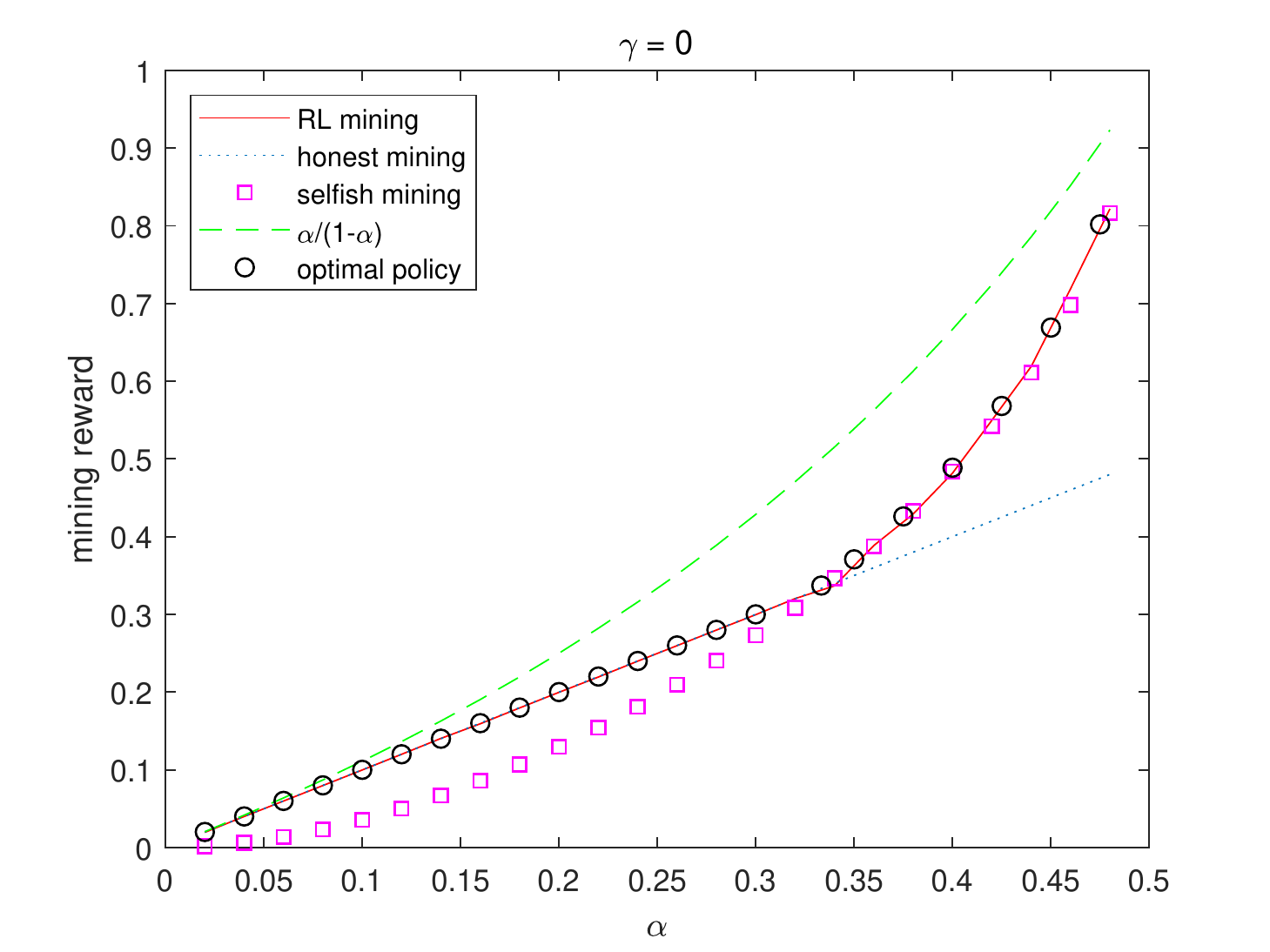}
	\caption{Achieved mining gain versus $\alpha $ for $\gamma =0$. } \label{fig_simu3}
\end{figure}

We first compare the performances of our RL mining, the optimal-policy mining,  and the selfish mining. Fig. \ref{fig_simu1}-\ref{fig_simu3} plots the mining reward of the adversary versus $\alpha $ for different values of $\gamma$ and $\gamma  \in \left\{ {0,0.5,1} \right\}$. {\color{black}{Note that the value of $\gamma $ ranges in the interval [0, 1]. Therefore, $\gamma=0$, $\gamma=0.5$, $\gamma=1$ respectively means a low, a median, and a high communication capability for the adversary. We just take these three values for $\gamma $ to demonstrate that the adversary with our RL mining can dynamically adapt to the optimal mining when it has different communication capability.}} The relative mining gain of  ${\alpha  \mathord{\left/
		{\vphantom {\alpha  {\left( {1 - \alpha } \right)}}} \right.
		\kern-\nulldelimiterspace} {\left( {1 - \alpha } \right)}}$ is treated as a bound for the mining problem and it can only be achieved by optimal-policy mining for $\gamma=1$. To derive the optimal policy, we adopt the search algorithm proposed in \cite{sapirshtein2016optimal} and set the search error to a very tiny number of  ${10^{ - 5}}$. As in \cite{sapirshtein2016optimal}, we truncate the MDP at  ${l^{\left( a \right)}} = 100$ or ${l^{\left( a \right)}} = 100$  for both of optimal-policy mining and RL mining. The temperature parameter $T_\varepsilon$ is set to as ${T_\varepsilon } = {10^4}$ and it is reset to  $T_\varepsilon=0$  after $t = {10^8}$ time steps when convergence is attained. All the results of RL mining are given after the algorithm has converged.  From the results, we can see that the performance of our RL mining can converge to the performance of optimal-policy mining without knowing the details about the environment model.

\begin{figure}[t]
	\centering
	\includegraphics[width=3.2in]{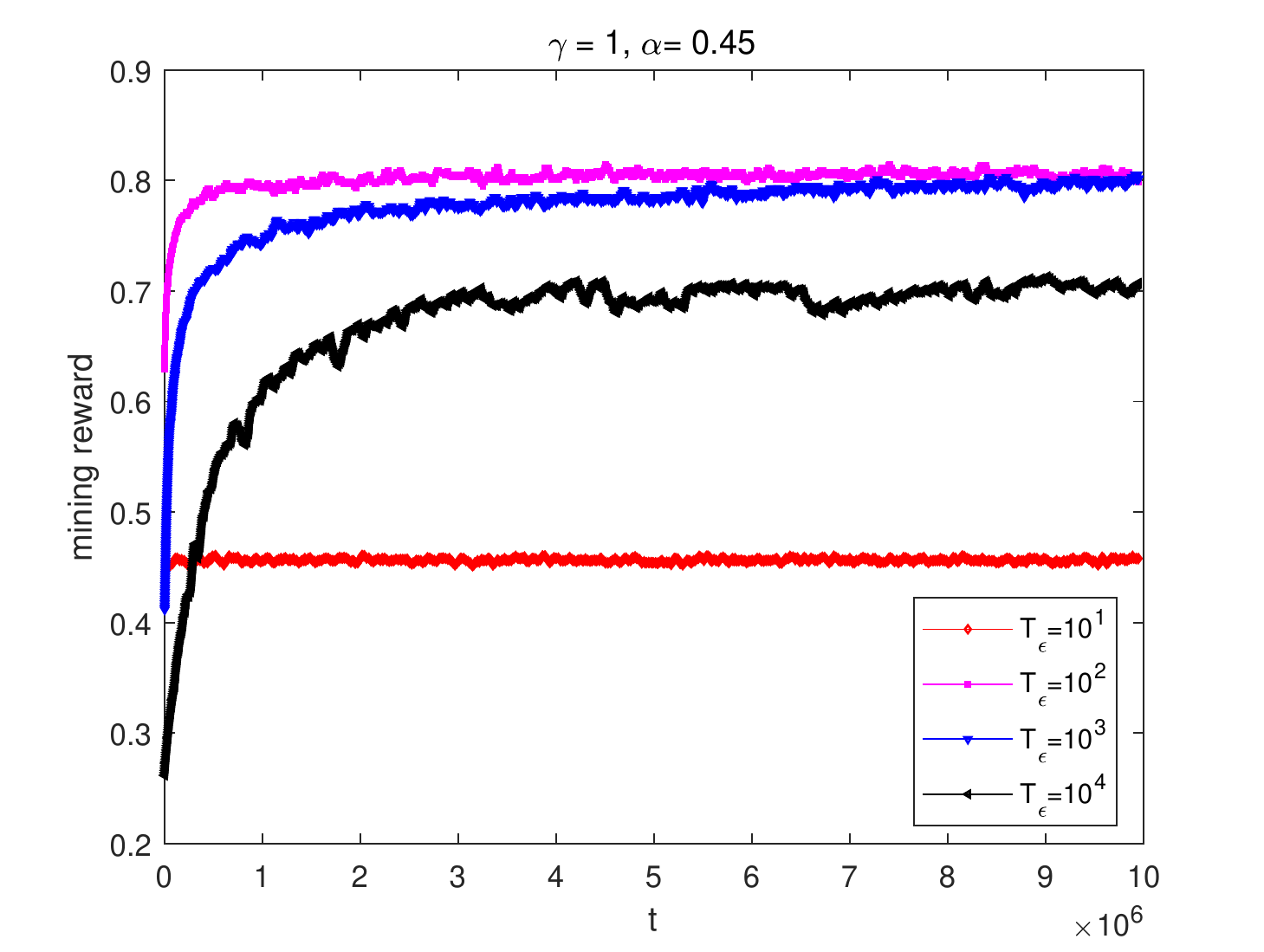}
	\caption{Achieved mining gain versus time step for different $T_\varepsilon$ and  $\gamma =1$, $\alpha=0.45$.} \label{fig_simu4}
\end{figure}

\begin{figure}[t]
	\centering
	\includegraphics[width=3.2in]{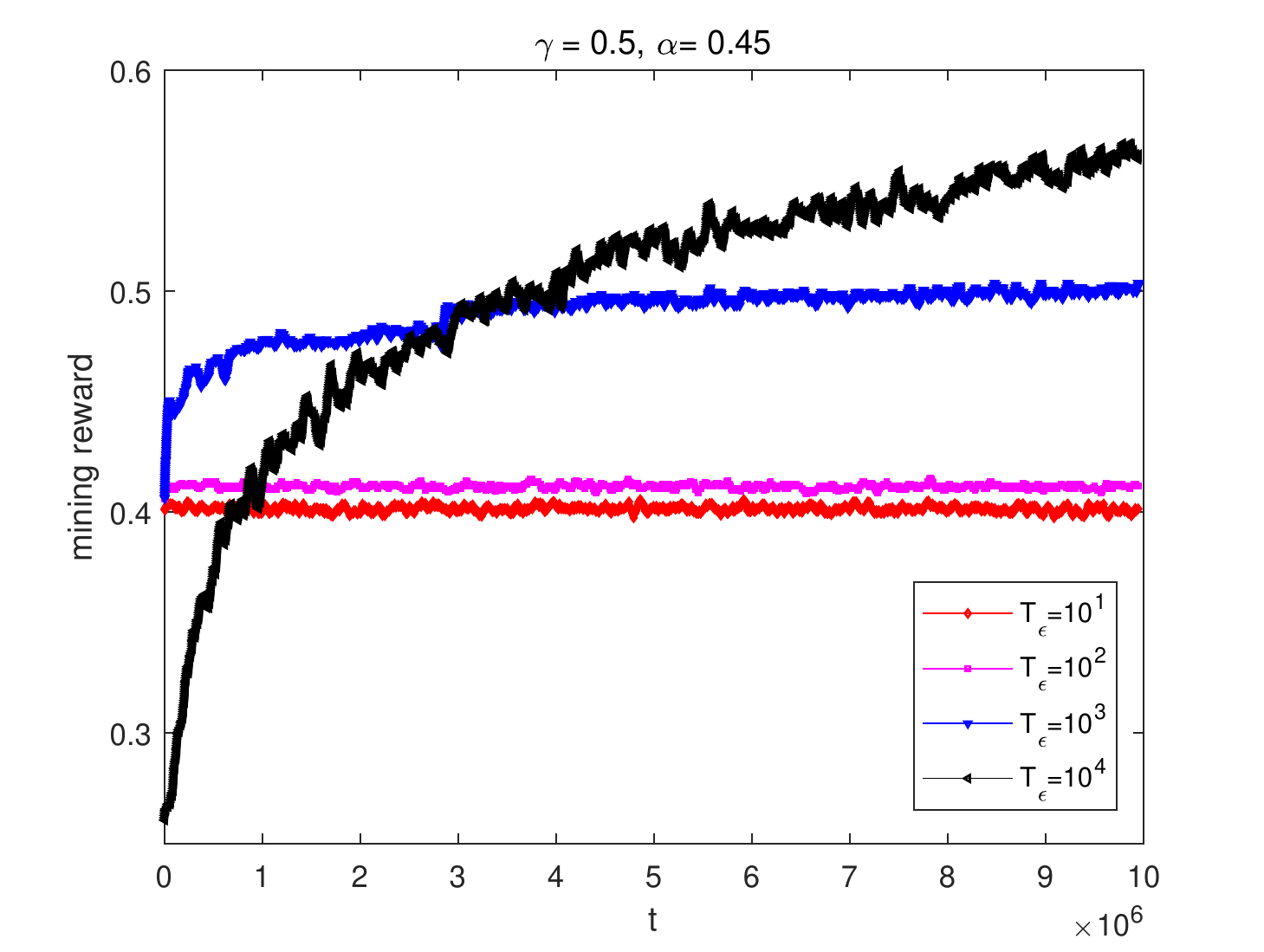}
	\caption{Achieved mining gain versus time step for different $T_\varepsilon$ and $\gamma =0.5$, $\alpha=0.45$.} \label{fig_simu5}
\end{figure}

\begin{figure}[t]
	\centering
	\includegraphics[width=3.2in]{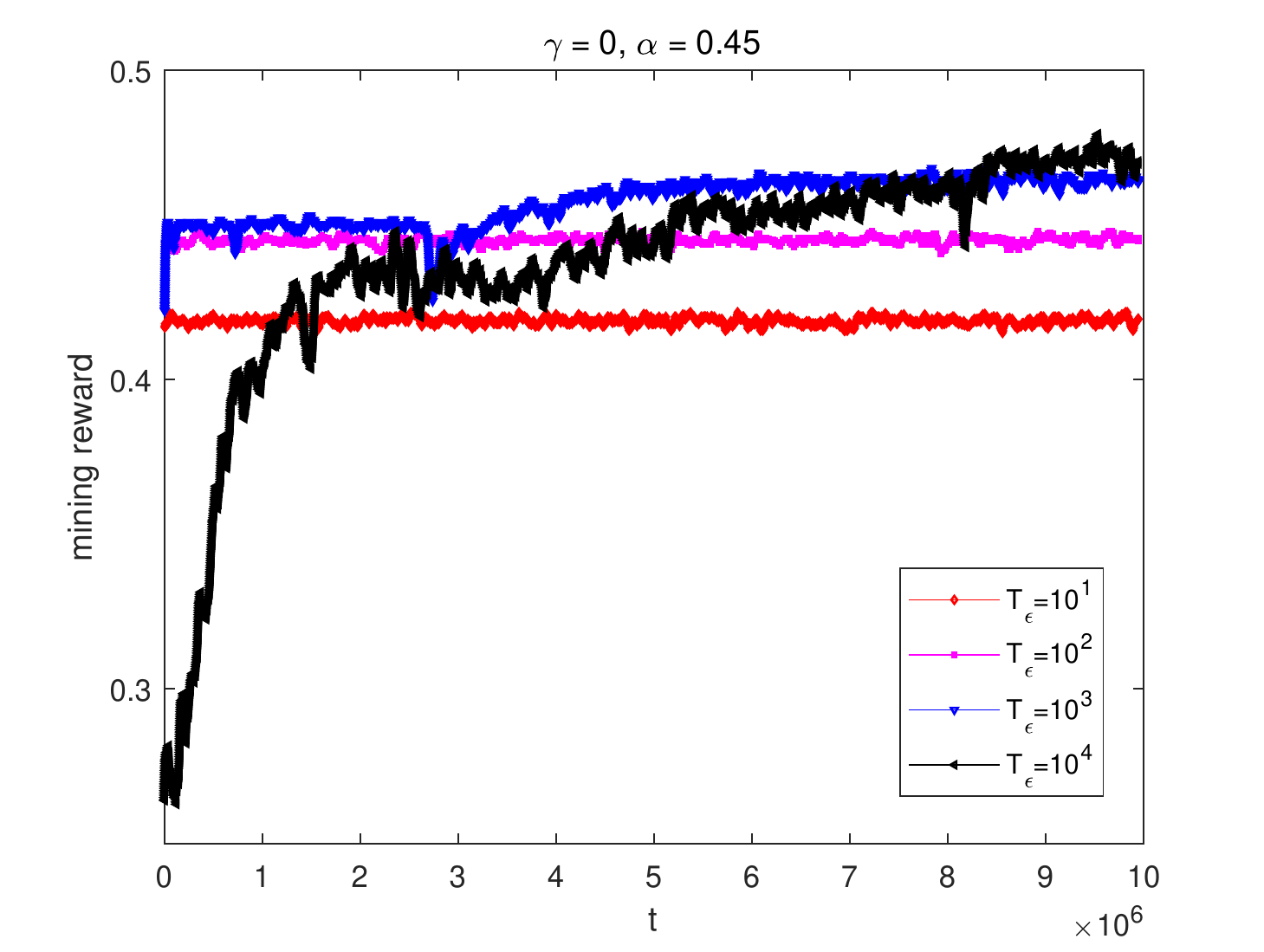}
	\caption{Achieved mining gain versus time step for different $T_\varepsilon$ and $\gamma =0$, $\alpha=0.45$.} \label{fig_simu6}
\end{figure}

\begin{figure}[t]
	\centering
	\includegraphics[width=3.5in]{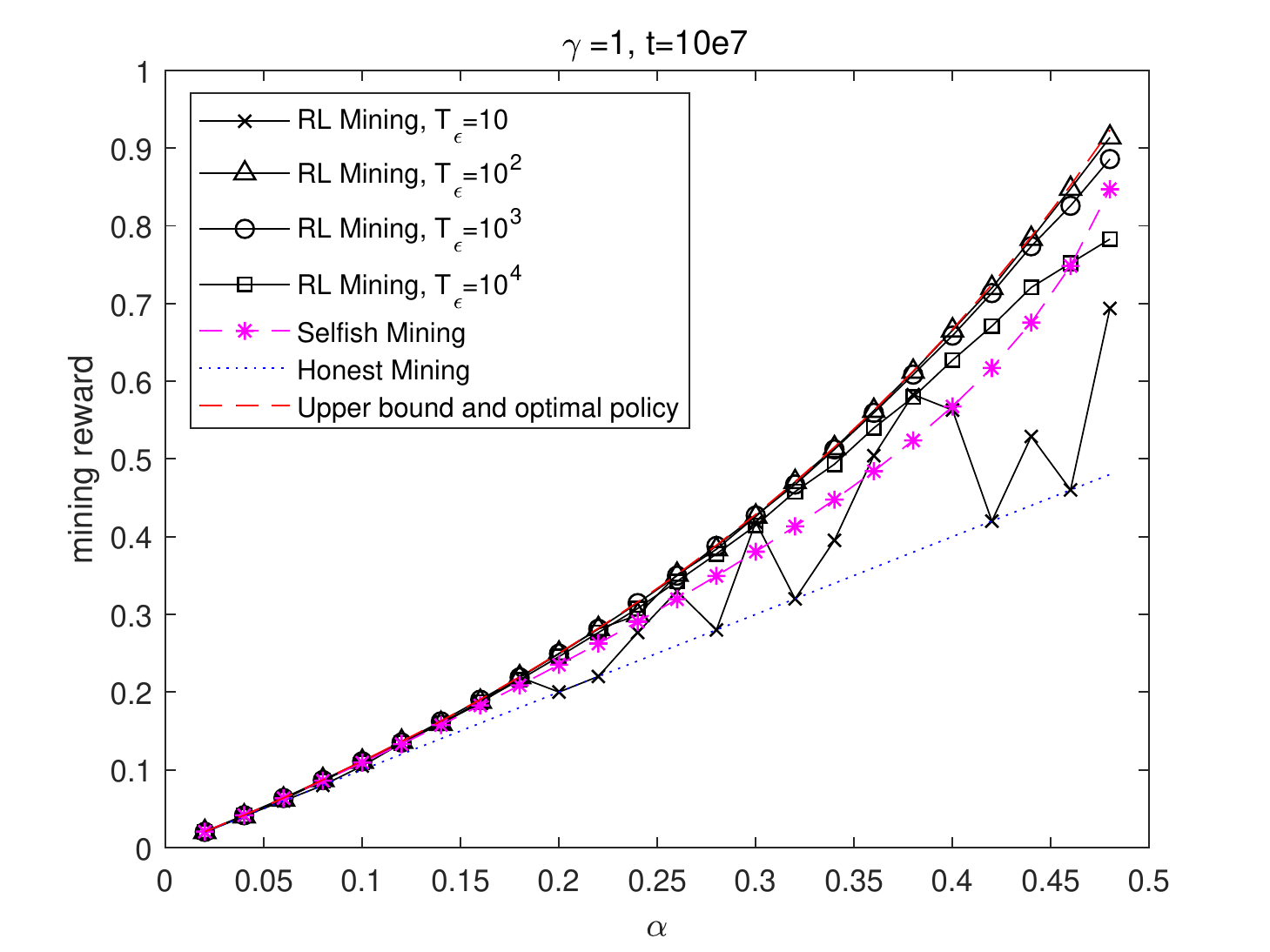}
	\caption{Achieved mining gain versus the $\alpha$ for $\gamma =1$ and different $T_\varepsilon$. } \label{fig_simu7}
\end{figure}

We next consider the impact of the temperature parameter $T_\varepsilon$  on the convergence of RL mining. Fig. \ref{fig_simu4}-\ref{fig_simu6} present the mining rewards obtained by RL mining with different $T_\varepsilon$  over time for $\gamma  \in \left\{ {0,0.5,1} \right\}$, respectively ($\alpha$ is fixed to 0.45). {\color{black}{In fact, the parameter of $T_{\epsilon}$ determines the extent of the exploration performed by the RL algorithm in its learning process. A larger $T_{\epsilon}$ encourages more explorations, and eventually, the learning process can converge, although a larger $T_{\epsilon}$ needs more time to converge. The optimal value of $T_{\epsilon}$ can lead to the convergence of RL by enough explorations and does not waste learning time by having unnecessary explorations. How to tune to the optimal $T_{\epsilon}$ is an interesting research direction. In this work, we just investigate the impact of $T_{\epsilon}$ on our RL mining by simulations.}} From the simulation results in Fig. \ref{fig_simu4}-\ref{fig_simu6}, we can see that generally, RL mining with larger $T_\varepsilon$  can have more explorations and can converge more closely to the optimal performance; however, RL mining with larger $T_\varepsilon$  also have longer exploration phases that slow down the convergence process. Fig. \ref{fig_simu7} presents the mining rewards of RL mining with different $T_\varepsilon$ for different $\alpha$ ($\gamma $ is fixed to 1). The mining reward results are given after $t = {10^7}$ time steps and without resetting $T_\varepsilon=0$. We see that for larger $\alpha$, we need larger $T_\varepsilon$  to ensure the convergence of RL mining, although it will slow down the convergence process. In practice, we can dynamically reduce the value of $T_\varepsilon$  when we find that the mining gain has already converged.

\begin{table}[t]
	\centering\centering
	\caption{The optimal policy for the blockchain environment with $\left( {\alpha  = 0.{\rm{35}},{\rm{ }}\gamma {\rm{ = 1}}} \right)$ when ${l^{\left( a \right)}} \le 8$ and ${l^{\left( h \right)}} \le 8$.}
	\begin{tabular}{lllllllll}
		\hline
		{${l^{\left( a \right)}}$}{${l^{\left( h \right)}}$} & 1   & 2   & 3   & 4   & 5   & 6   & 7   & 8   \\ \hline
		1   & *** & *a* & *** & *** & *** & *** & *** & *** \\
		2   & w** & *m* & *w* & *a* & ***    & ***    & ***     & ***    \\
		3   & w** & *oo & w** & *w* & *a*    & ***    & ***     & ***     \\
		4   & w** & *m* & oo* & w** & *w*     & *a*    & ***     & ***    \\
		5   & w** & *mw & *m* & oo* & w**    &  *w*    & *w*      & *a*    \\
		6   & w** & *mw & *mw & *m* & oo*    & w**    & ww*      & *w*    \\
		7   & w** & *mw & *mw & *mw & *m*    & oo*    & w**     & ww*    \\
		8   & w** & *mw & *mw & *mw & *mw    & *m*    & oo*    & w**  	\\ \hline
	\end{tabular}
\end{table}

\begin{figure}[t]
	\centering
	\includegraphics[width=3.2in]{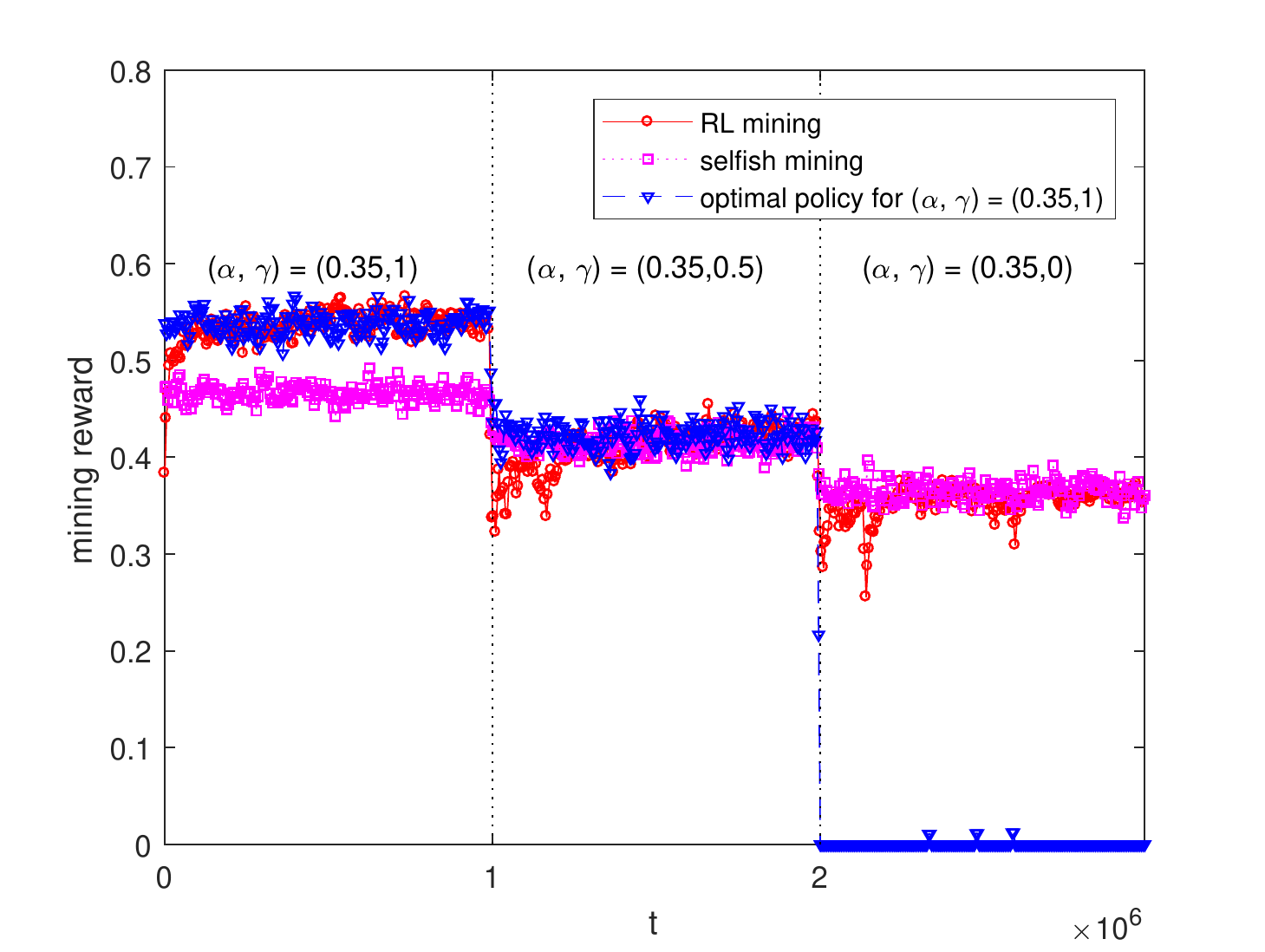}
	\caption{Achieved mining gain when the environment is changing and the values  of  $\left( {\alpha ,\gamma } \right)$ change  in  the  following  order:  (0.35,1),  (0.35, 0.5),  and  (0.35,0).} \label{simu_changing_env1}
\end{figure}

\begin{figure}[t]
	\centering
	\includegraphics[width=3.2in]{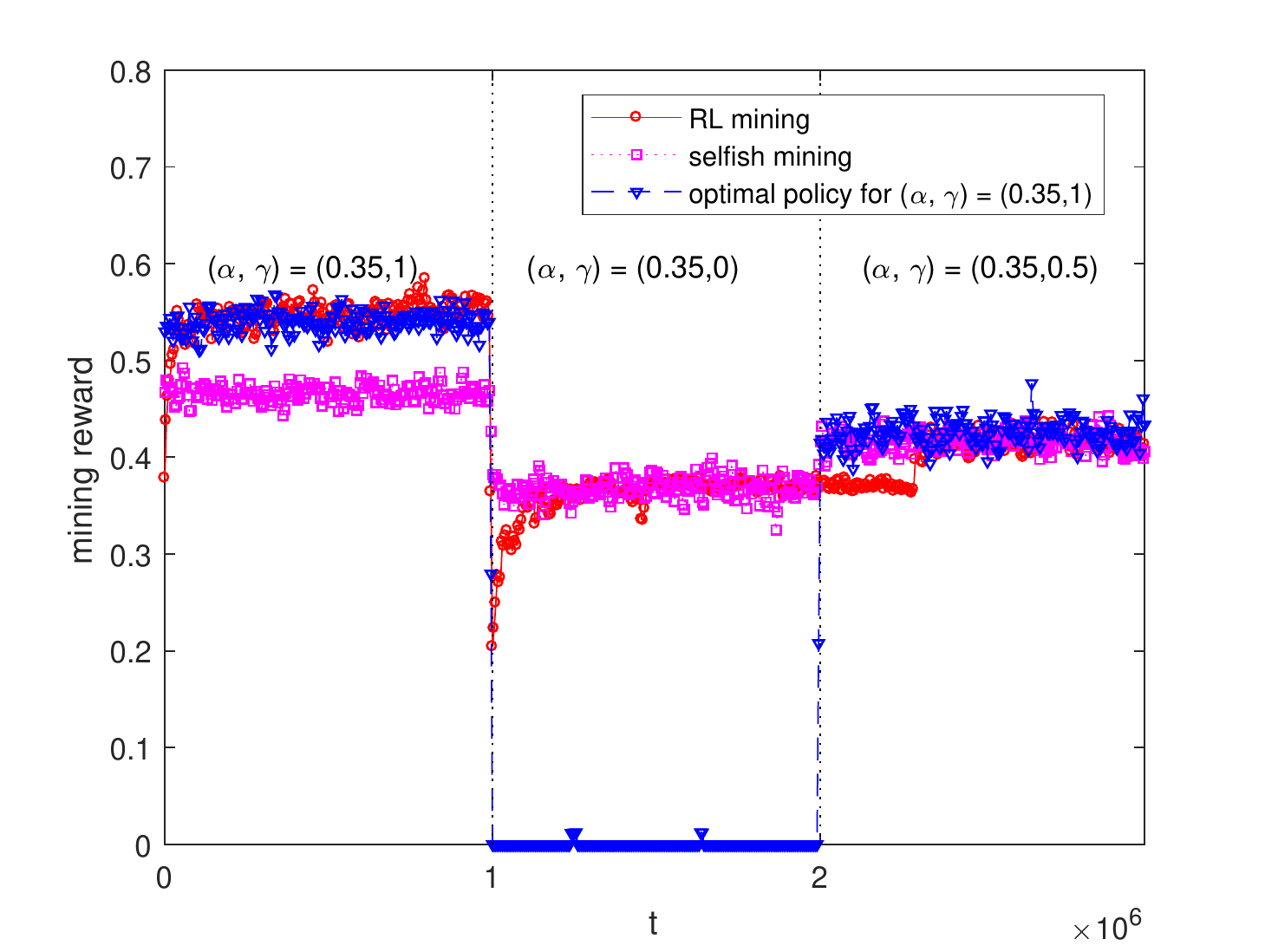}
	\caption{Achieved mining gain when the environment is changing and the values  of  $\left( {\alpha ,\gamma } \right)$ change  in  the  following  order:  (0.35,1),  (0.35,  0),  and  (0.35,0.5).} \label{simu_changing_env2}
\end{figure}

\begin{figure}[t]
	\centering
	\includegraphics[width=3.2in]{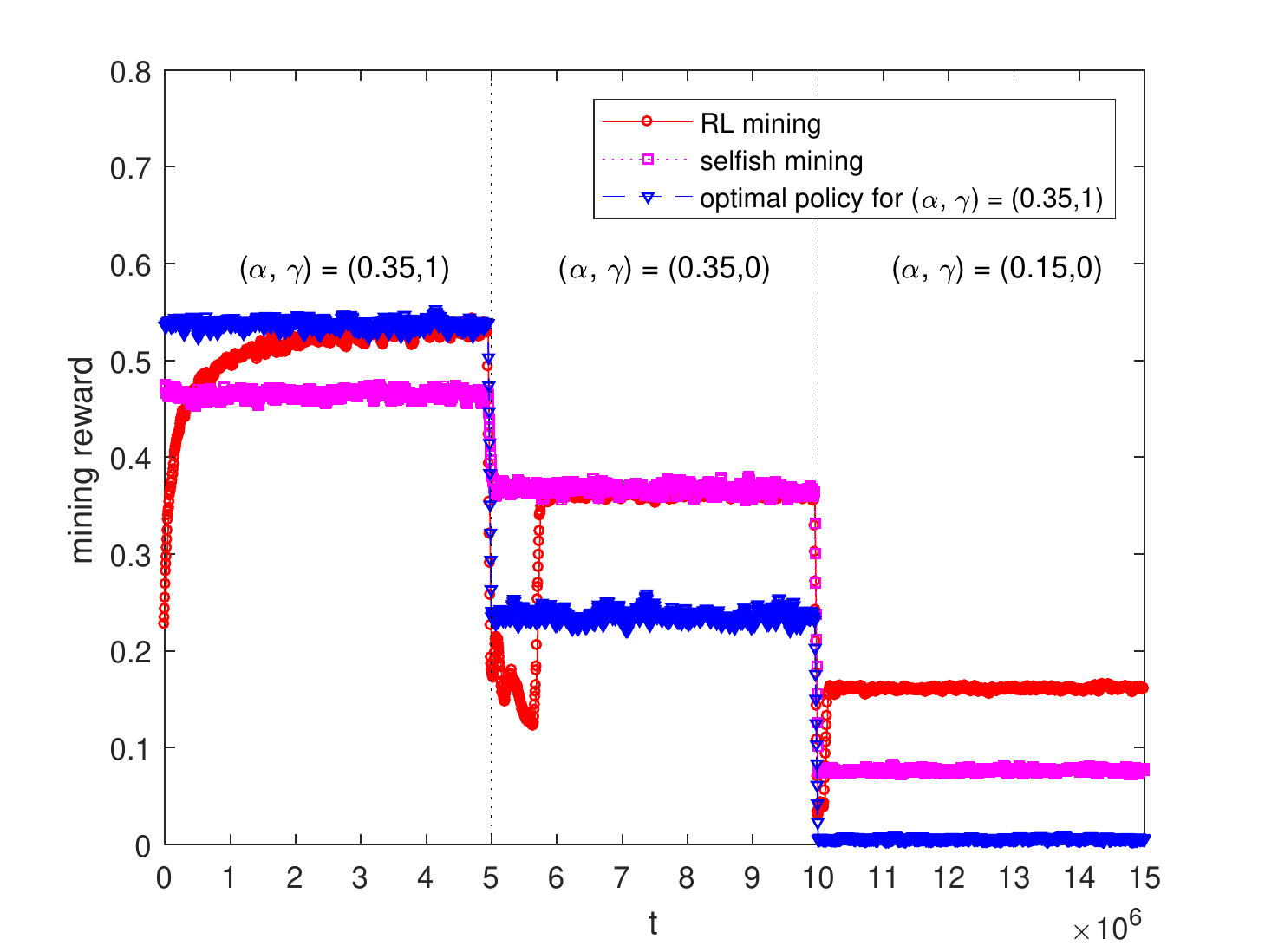}
	\caption{Achieved mining gain when the environment is changing and the values  of  $\left( {\alpha ,\gamma } \right)$ change  in  the  following  order:  (0.35,1),  (0.35,  0),  and  (0.15,0).} \label{simu_changing_env3}
\end{figure}

Last, we investigate the mining performances of different mining strategies when the blockchain environment changes. The experimental results are given in Fig. \ref{simu_changing_env1}-\ref{simu_changing_env3}. The blockchain environment starts with parameter values of $\left( {\alpha  = 0.{\rm{35}},{\rm{ }}\gamma {\rm{ = 1}}} \right)$ and the values of $\left( {\alpha ,\gamma } \right)$ change sequentially in the experiment. The temperature parameter  $T_\varepsilon$ of RL mining is fixed to ${T_\varepsilon } = {10^3}$. The optimal-policy mining strategy adopts the optimal policy for the blockchain environment with $\left( {\alpha  = 0.{\rm{35}},{\rm{ }}\gamma {\rm{ = 1}}} \right)$. We derived the optimal policy for $\left( {\alpha  = 0.{\rm{35}},{\rm{ }}\gamma {\rm{ = 1}}} \right)$ by iteratively exploit the MDP solver \cite{chades2014mdptoolbox} to search over the policy space, as proposed in \cite{sapirshtein2016optimal}. TABLE II describes the found optimal policy for $\left( {\alpha  = 0.{\rm{35}},{\rm{ }}\gamma {\rm{ = 1}}} \right)$ when ${l^{\left( a \right)}} \le 8$ and ${l^{\left( h \right)}} \le 8$. The performances of the optimal policy for $\left( {\alpha  = 0.{\rm{35}},{\rm{ }}\gamma {\rm{ = 1}}} \right)$, and the selfish mining are treated as benchmarks for our RL mining in the changing blockchain environment. {\color{black}{In Fig. 16-18, for different values of the parameters $(\alpha, \gamma) $, the performances of optimal selfish mining are still obtained using the policy of the optimal selfish mining under model parameters $(\alpha, \gamma) = (0.35,1)$.}} From the simulation results, we can see that when the environment has changed, the optimal-policy mining strategy derived from the MDP model is not optimal anymore; our RL mining can adaptively learn the optimal policy for different environments. This demonstrates the advantage of RL mining over these model-based mining strategies.

\section{Conclusion}

We employed RL algorithms to solve the mining MDP problem of Bitcoin-like blockchains. We showed that, without knowing parameters about the blockchain network model, our RL mining can achieve the mining reward of the optimal policy that requires knowledge of the parameters. Therefore, in a dynamic environment in which the parameter values can change over time, RL mining can be more robust.  

Going forward, we will investigate two issues that need to be addressed before RL mining can be practical:  

1.	More complete MDP model as proposed in \cite{gervais2016security} for blockchain networks---This model incorporates detailed blockchain features, such as stale block rate, double spending attack, and eclipsed attack, that have been precluded by the model in the current paper.  The large action-space of the complete  model will make it more challenging for RL mining to learn an optimal strategy.    

2.	Cost of lagged time in convergence---Since miners need to pay for their hardware and consume electricity to mine blocks, fast convergence of the mining algorithm is important from the economical standpoint. Deep RL \cite{mnih2015human} that incorporates deep neural networks into RL can potentially speed up the convergence rate. We will consider the exploit of deep RL in our future work.

\ifCLASSOPTIONcaptionsoff
  \newpage
\fi

\bibliographystyle{IEEEtran}

\bibliography{refs}

\end{document}